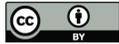

**Hydrology and Earth System Sciences**

Open Access

# Early 21st century snow cover state over the western river basins of the Indus River system

S. Hasson[1,2], V. Lucarini[1,4], M. R. Khan[3], M. Petitta[5], T. Bolch[6,7], and G. Gioli[2]

[1]CEN, Centre for Earth System Research and Sustainability, Meteorological Institute, University of Hamburg, Hamburg, Germany
[2]CEN, Centre for Earth System Research and Sustainability, Institute for Geography, University of Hamburg, Hamburg, Germany
[3]Department of Geo-informatics, PMAS-Arid Agriculture University, Rawalpindi, Pakistan
[4]Department of Mathematics and Statistics, University of Reading, Reading, UK
[5]Institute for Applied Remote Sensing, EURAC, Bolzano/Bozen, Italy
[6]Department of Geography, University of Zurich, Zurich, Switzerland
[7]Institute for Cartography, Technische Universität Dresden, Dresden, Germany

*Correspondence to:* S. Hasson (shabeh.hasson@zmaw.de)

**Abstract.** In this paper we assess the snow cover and its dynamics for the western river basins of the Indus River system (IRS) and their sub-basins located in Afghanistan, China, India and Pakistan for the period 2001–2012. First, we validate the Moderate Resolution Imaging Spectroradiometer (MODIS) daily snow products from Terra (MOD10A1) and Aqua (MYD10A1) against the Landsat Thematic Mapper/Enhanced Thematic Mapper plus (TM/ETM+) data set, and then improve them for clouds by applying a validated non-spectral cloud removal technique. The improved snow product has been analysed on a seasonal and annual basis against different topographic parameters (aspect, elevation and slope). Our results show a decreasing tendency for the annual average snow cover for the westerlies-influenced basins (upper Indus basin (UIB), Astore, Hunza, Shigar and Shyok) and an increasing tendency for the monsoon-influenced basins (Jhelum, Kabul, Swat and Gilgit). Seasonal average snow cover decreases during winter and autumn, and increases during spring and summer, which is consistent with the observed cooling and warming trends during the respective seasons. Sub-basins at relatively higher latitudes/altitudes show higher variability than basins at lower latitudes/middle altitudes. Northeastern and northwestern aspects feature greater snow cover. The mean end-of-summer regional snow line altitude (SLA) zones range from 3000 to 5000 m a.s.l. for all basins. Our analysis provides an indication of a descending end-of-summer regional SLA zone for most of the studied basins, which is significant for the Shyok and Kabul basins, thus indicating a change in their water resources. Such results are consistent with the observed hydro-climatic data, recently collected local perceptions and glacier mass balances for the investigated period within the UIB. Moreover, our analysis shows a significant correlation between winter season snow cover and the North Atlantic Oscillation (NAO) index of the previous autumn. Similarly, the inter-annual variability of spring season snow cover and spring season precipitation explains well the inter-annual variability of the summer season discharge from most of the basins. These findings indicate some potential for the seasonal stream flow forecast in the region, suggesting snow cover as a possible predictor.

## 1 Introduction

Snow is an essential part of the climate system, with a large influence on the hydrological cycle as well as on the atmospheric processes due to its high albedo and low thermal conductivity (Hall and Riggs, 2007). Snow is especially important for the hydrological cycle, as large amounts of the water supplies come from seasonal snowmelt at high latitudes and in mountainous basins (Barnett et al., 2005). This is particularly true for the Indus basin, where snowmelt from the Hindu Kush–Karakoram–Himalaya (HKH) ranges provides the first water available after a long dry period (October–March, Immerzeel et al., 2010) that is mainly used for irrigation and power generation (Hasson et al., 2013). Besides its importance, the snowmelt contribution to the Indus flows is not well known. Immerzeel et al. (2009) have reported the snowmelt contribution to the upper Indus basin (UIB) flows to be around 40 %. Such an estimate is based on a modelling study and is subject to various uncertainties associated with the modelling approaches. Nevertheless, in view of the importance of snowmelt for the Indus basin, an accurate quantification of snow distribution patterns and their climatic properties is essential. Additionally, snow cover assessment is required for the calibration/validation of distributed hydrological models (Konz et al., 2010), and for the seasonal forecast of freshwater supplies.

Consistent with the unprecedented global warming (IPCC, 2013), the Himalayas have experienced significant warming during recent decades (Shrestha et al., 1999; Diodato et al., 2012). Consequently, the annual average snow cover decreased by ∼ 16 % over the entire Himalayas from 1990 to 2001 (Menon et al., 2010). A similar snow cover trend has also been observed from 2000 to 2008 (Immerzeel et al., 2009), though some regional anomalies do exist. Similarly, most of the Himalayan glaciers have been retreating and losing mass since the end of the Little Ice Age, where current observations show, on average, an acceleration of such responses since the mid-1990s (Bolch et al., 2012). In contrast, the UIB experiences unique signatures of climate change, featuring cooling temperatures and increasing precipitation (Fowler and Archer, 2006). Tree-ring-based precipitation reconstruction further confirms that the last century was the wettest in the last millennium in this region (Treydte et al., 2006). The Karakoram glaciers within the UIB have featured irregular behaviour for a long time, showing balanced budgets and several advancing glaciers, especially during the last decade (Hewitt, 2005; Bolch et al., 2012; Bhambri et al., 2013; Gardelle et al., 2013). Local narratives also confirm the anomalous features of hydro-climatic changes within the UIB (Gioli et al., 2014). Under such contrasting hydro-climatic regimes, the prevailing snow cover state is largely unknown, leading to uncertainties in the present and future management of water resources.

The sparse network of short-length high-altitude meteorological stations within Pakistan makes it hard to assess the detailed picture of snow cover dynamics on sub-basin and regional scales. Regional snow surveys are also not possible in the HKH region, due to its complex terrain and harsh environment. Furthermore, as snow features a high degree of variability, it needs mapping at a fine temporal resolution, unlike glaciers, which require a fine planar resolution. In this regard, the integration of remote sensing (RS) data and methods with geographical information system (GIS) techniques has proved its usefulness in mapping snow cover in inaccessible areas (Max, 2001; Tong et al., 2009). Moreover, satellite sensors offer a unique opportunity for snow cover monitoring, supporting the efficient management of the snow-covered areas (Notarnicola et al., 2011; Thirel et al., 2012). Some examples are the Moderate Resolution Imaging Spectroradiometer (MODIS) sensor on-board NASA's Earth Observing System (EOS) Aqua and Terra satellites (NASA MODIS), the Advanced Very High Resolution Radiometer (AVHRR on the NOAA-15 satellite), or the Interactive Multisensor Snow and Ice Mapping System – IMS, which incorporates a variety of satellite images from AVHRR, GOES, and SSMI (Helfrich et al., 2007; Roy et al., 2010). However, remote sensing of snow is subject to various limitations, such as obstruction of snow by dense vegetation, surface heterogeneity, spectral similarities between different objects in mountainous areas, and cloud cover impeding the surface view (López-Burgos et al., 2013; Hüsler et al., 2014). Dietz et al. (2012) provide a detailed review of the common methods of snow cover mapping from the remotely sensed data, along with their limitations and advantages. Unfortunately, satellite data have a quite limited ability in retrieving any direct information about the snow amount, such as snow depth (SD) and snow water equivalent (SWE). Presently, available satellite-based SD/SWE observations show large discrepancies (Dong et al., 2005; Takala et al., 2011), and their coarse resolution of around 25 km is unsuitable for mountainous areas. This hinders an adequate analysis of the snowpack water content in the region.

Recent studies have presented snow cover estimates for Hunza, Astore and UIB (Immerzeel et al., 2009; Tahir et al., 2011a; Forsythe et al., 2012). Such estimates are based on the MODIS 8-day maximum snow cover product, which is biased positive (Xie et al., 2009) and features cloud data gaps. As clouds are one of the major problems for optical remote sensing data, their presence in MODIS snow products prevents adequate snow assessment and introduces uncertainties in the analysis (Hall et al., 2002). This emphasises the need to improve considered snow products by filling up cloud data gaps prior to their use in the analysis. Since the MODIS daily snow product features a larger period of high snow classification agreement between Terra and Aqua (Wang et al., 2009), it is highly suitable for filling cloud data gaps. Furthermore, large-scale snow accumulation/distribution processes are generally controlled by synoptic-scale meteorological patterns and the large-scale topographical properties (especially elevation and aspect). Instead, snow redistribution depends on the local topography (especially slope) and local meteorological conditions on a small scale (Kelly et al., 2003). In the present study, we first validate the hyper-temporal MODIS daily snow products from Terra (MOD10A1) and Aqua (MYD10A1) against Landsat TM (Thematic Mapper)/ETM+ (Enhanced Thematic Mapper plus) images, and then improve the validated snow product for clouds by applying a validated rigorous non-spectral cloud removal technique. Selecting practical study domains in view of melt-runoff modelling and water resource management, we then present improved snow cover estimates for the western river basins (Indus, Jhelum and Kabul) and their six sub-basins against different topographic parameters (aspect, elevation and slope). In view of scarce SD/SWE observations, for the first time we present a comprehensive picture of precipitation input and its distribution over the main study region up to an altitude of 4500 m a.s.l. from 36 meteorological stations. We also report the precipitation estimates either collected during the short-term field campaigns or from the stations not studied here. Furthermore, we provide proxy evidence of a qualitative change in the mass balance of existing glaciers by ascertaining end-of-summer regional SLA zone tendencies. We successfully link such findings to recent hydro-climatic signals as well as to the socio-economic vulnerability observed over recent decades. In addition, we investigate whether the North Atlantic Oscillation (NAO) can explain the snow cover variability over the region, and whether such variability further contributes to explaining the variability of melt-season runoff with a reasonable lead time, in order to explore the possibility of a runoff forecast well in advance for better management of water resources in the region.

## 2 Study area

The study area covers the spatial domain of 30–38° N and 67–84° E and encompasses large part of the HKH ranges. It comprises of three large trans-boundary river basins, such as, Jhelum, Kabul and UIB and their six sub-basins, namely Astore, Gilgit, Hunza, Swat, Shigar and Shyok (Fig. 1), located in Afghanistan, China, India and Pakistan. According to the Indus water treaty (1960), Indus River and two out of its five eastern tributaries, such as Jhelum and Chenab, are called the western rivers of the Indus River System (IRS) (Mehta, 1988). Here, we consider only basins of Indus and Jhelum rivers but include a basin of the Kabul River, which is a western tributary of the Indus River. The investigated river basins are located at the boundary between two large-scale circulation modes: the western mid-latitude disturbances and the south Asian summer monsoon system (Hasson et al., 2014). The hydrology of the high latitude/altitude (Hunza, Shigar, Shyok and Astore) sub-basins is dominated by the precipitation regime of the western mid-latitude disturbances during the winter and spring seasons (Wake, 1987; Rees and Collins, 2006; Hewitt, 2011; Hasson et al., 2013). Such high latitude/altitude sub-basins, located in the rain shadow of the western Himalayas, receive negligible precipitation from the summer monsoon (Ali et al., 2009), which is mostly restricted to the lower latitude/altitude (Swat, Jhelum, Kabul and Gilgit) sub-/basins. Runoff from the study basins primarily depends upon the slow (snow and glacier melt) and a fast (rainfall) component in the higher and lower altitude sectors, respectively (Archer, 2003; Ali and De Boer, 2007; Hasson et al., 2013). It is confined to the summer months (June–September) and provides almost 80 % of the annual surface water available within Pakistan (Ali et al., 2009). Based on the hydro-meteorological characteristics of the study region, Fowler and Archer (2005) have divided it into three major categories:

– High altitude glacier-fed basins with a large percentage of glacier cover, whose runoff primarily depends upon the glacier melt and strongly correlates with the concurrent summer temperatures. The snow distribution significantly affects the timing and magnitude of the glacier melt runoff from these basins.

– Mid-altitude snow-fed basins with lower elevation and smaller percentage of glacier cover than the glacier-fed basins, whose runoff primarily depends upon the snowmelt and strongly correlates with the previous winter season solid precipitation.

– Low-altitude foothill rain-fed basins, which mostly receive precipitation in a liquid form.

Though dominated by the slow runoff component, the hydrology of the snow- and glacier-fed basins can further be differentiated on the basis of their runoff production time – e.g. the peak runoff during June and August, respectively. Therefore, we have considered the glacier-fed (Hunza, Shigar, and Shyok) and snow-fed (Jhelum, Kabul, Gilgit, Astore, and Swat) basins separately for our further investigations (Fig. 1, Table 1).

The considered basins feature conflicting signals of climate change. For instance, almost half of the observational record within the UIB shows a cooling tendency of the mean annual and seasonal temperatures, except during the winter season since the 1960s (Fowler and Archer, 2006). Consequently, the melt season runoff is declining (Khattak et al., 2011). Furthermore, the diurnal temperature range is widening in the UIB throughout the year (Fowler and Archer, 2006), while it has been narrowing worldwide since 1950 (Karl et al., 1993; Easterling et al., 1997). However, a statistically significant increase for the summer, winter and annual precipitation has been observed during the second half of the 20th century (Archer and Fowler, 2004). Such hydro-meteorological phenomena along with subsequent heterogeneous responses from the existing cryosphere determine the overall hydrological balance of the study basins and the water availability downstream in an otherwise very arid land.

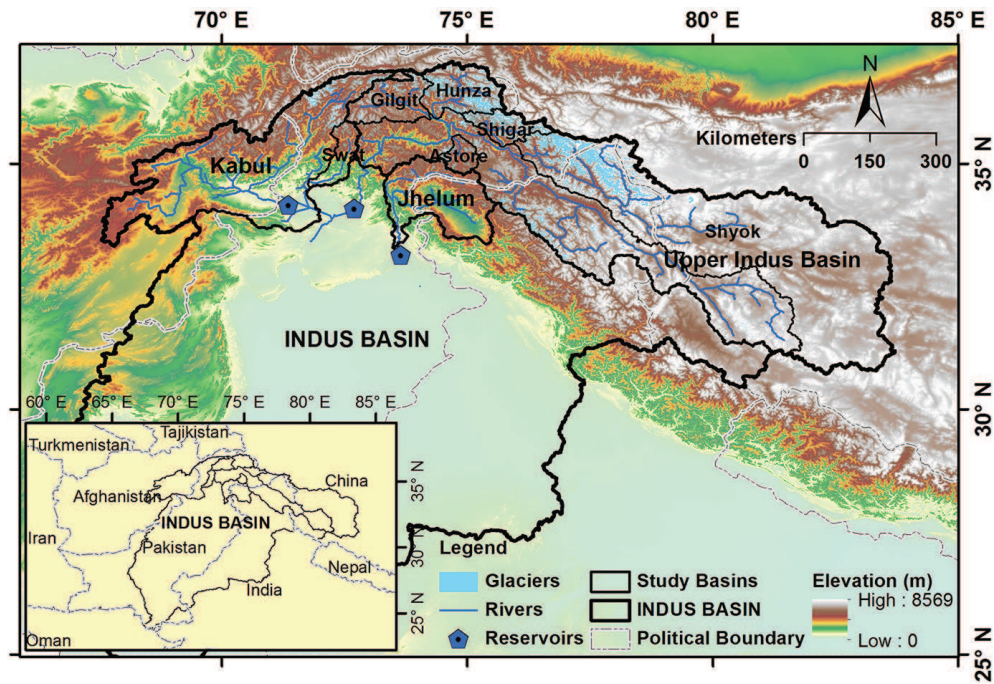

**Figure 1.** Study area for snow cover mapping showing three major western river basins, namely Indus, Kabul and Jhelum and their six sub-basins. Swat is a sub-basin of the Kabul basin, whereas Gilgit, Hunza, Shigar, Shyok and Astore are the sub-basins of UIB. The political boundaries are tentative.

**Table 1.** Study basins, their areas and glacier coverage. Note: glacier areas are derived from the Randolph Glacier Inventory (Pfeffer et al., 2014).

| S. no. | Basin at gauging site | Area (km$^2$) | Glacier area (%) |
|---|---|---|---|
| 1 | Astore at Doyian | 3897 | 14 |
| 2 | Gilgit at Gilgit | 12 652 | 7 |
| 3 | Hunza at Dainyor Bridge | 13 705 | 28 |
| 4 | Jhelum at Azad Patan | 27 291 | 1 |
| 5 | Kabul at Nowshera | 88 676 | 2 |
| 6 | Swat at Chakdara | 6080 | 3 |
| 7 | Shigar at Shigar | 6974 | 30 |
| 8 | Shyok at Yugo | 138 836 | 6 |
| 9 | UIB at Besham Qila | 271 359 | 7 |

## 3 Data

### 3.1 Snow cover data

The MODIS sensors on-board both the Terra and Aqua satellites, passing over the same area in the morning and afternoon, respectively, provide imagery with twice a daily temporal resolution. A normalised difference snow index (NDSI) along with the normalised difference vegetation index (NDVI) is used to detect snow from the MODIS imagery through an automated procedure. Snow products are then produced ranging from a swath level to a composite global climate modelling grid (CMG) after spatial and temporal transformations. Each higher level snow product, therefore, assimilates accuracy and errors from its preceding product (Riggs and Hall, 2011). In version 5 of the snow products, a surface temperature filtering has been applied to prevent the erroneous mapping of the warm surfaces as snow due to their spectral similarity (Riggs et al., 2006).

We use the MODIS daily snow products of 500 m resolution from both Terra and Aqua (MOD- and MYD-10A1) version 5 (Hall et al., 2006) for the period 2001–2012 for our analysis. The spatial resolution of 500 m has been considered highly suitable for estimating the snow cover of the basins with an area of about 10 000 km$^2$ or larger (Hall et al., 2002); however, it can still be useful for relatively smaller basins. In order to cover the study area, we have downloaded snow tiles of h23v5, h24v5 and h25v5 for both MODIS Terra and Aqua from the online archive of the NASA Distributed Active Archive Centre (DAAC) located at the National Snow and Ice Data Centre (NSIDC).

### 3.2 Satellite images

In view of scarce observations, previous studies have validated the MODIS snow products over part of our study area based on a proxy data set. For instance, Tahir et al. (2011a) have validated the 8-daily MODIS Terra snow product for the Hunza basin using images from Advanced Spaceborne Thermal Emission and Reflection Radiometer (ASTER). Forsythe et al. (2012) have validated the 8-daily MODIS Terra snow product for the Astore basin against the snow cover reconstructed from temperature records. Therefore, we have decided to validate the considered MODIS snow products against the Landsat images. We have downloaded 14 Landsat 5 TM and Landsat 7 ETM+ level 1T – terrain corrected – images from the Glovis online archive (www.glovis.usgs.gov). Landsat 7 ETM+ scenes are collected from 2001 to 31 May 2003 because scenes collected after this date feature large data gaps due to failure of the scan line corrector (SLC). However, Landsat 5 TM scenes are collected within the period 2001–2012, subject to their availability and clear sky conditions. Most of the collected images (TM/ETM+) are either cloud-free or observe cloud cover around 2 %. These images cover most of the maximum snow cover extent of the study area (Fig. 2), and belong to both accumulation and ablation seasons. Details of the Landsat 5 and 7 sensors and the images used here are given in the Supplement.

### 3.3 Digital elevation model (DEM)

The gap-filled Shuttle Radar Topography Mission (SRTM) digital elevation model (DEM) version 4 (Jarvis et al., 2008) at 90 m resolution, from the CGIAR Consortium for Spatial Information (http://srtm.csi.cgiar.org/), was used to define the topography and to delineate the watershed boundaries. It was interpolated to the MODIS snow product resolution of 500 m using the nearest neighbour method for the calculation of the required topographic parameters.

### 3.4 Hydro-meteorological data

For meteorological observations, we have obtained the SWE and precipitation data from all available sites within the Pakistan region of the study area. In 1995, the snow pillows were installed at various sites in the HKH region of Pakistan during the second phase of the Snow and Ice Hydrology Project (SIHP) of the Water and Power Development Authority (WAPDA), Pakistan through a joint venture with the Canadian team (SIHP, 1997). Most of the installed snow pillows, however, have so far faced technical issues of interfacing with the transmission system as well as unexpected "jumps" due to possible ice bridging and rupture effects (SIHP, 1997). Therefore, presently, the SWE observation from the Deosai (from the 2007–2008 period onwards) and Shogran (from the 2012–2013 period onwards) sites are available, offering only 5-year long time series at the Deosai station (2008–2012) within the analysis period (Fig. 3).

We have obtained the precipitation and temperature data from 18 high-altitude (between 2200 and 4500 m a.s.l.) SIHP stations within the HKH region for the period 1995–2012. These high altitude gauges measure both solid and liquid precipitation in mm water equivalent. Such capability of these high altitude precipitation gauges has allowed us to accumulate precipitation for the duration when temperatures contiguously remained below zero, in order to roughly estimate the winter/spring time SWE. However, such estimates may underestimate the actual amount of solid precipitation due to under-catch errors of the precipitation gauges under strong wind conditions. We have also collected the long-term precipitation record from 18 stations of relatively lower elevation (between 600 and 2200 m a.s.l.) for the period 1961–2010, which are being maintained by the Pakistan Meteorological Department (PMD), Pakistan. The discharge data have been obtained from the Surface Water Hydrology Project (SWHP) of WAPDA, Pakistan for all the nine study basins. The details of the collected hydro-meteorological data are given in the Supplement.

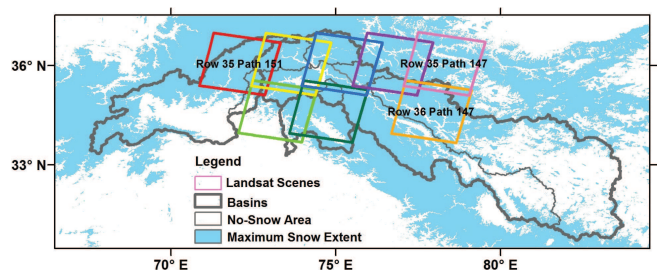

**Figure 2.** Landsat scenes used for validation of the MODIS Terra and Aqua products, covering most of the maximum snow cover extent of the studied basins. Note: each colour indicates a unique location ("Path" and "Row") of the Landsat scene acquisition.

### 3.5 NAO index

Additionally, the station-based seasonal mean North Atlantic Oscillation (NAO) index (Hurrell, 1995) was downloaded from an online archive (https://climatedataguide.ucar.edu/sites/default/files/climate_index_files/nao_station_seasonal.ascii) of the Climate Analysis Section, National Center for Atmospheric Research (NCAR), Boulder, USA, to analyse the statistical relation between snow cover and intensity and the position of the storm track and the ensuing mid-latitude disturbances. The station-based NAO index is calculated from the normalised sea level pressure differences between Ponta Delgada, Azores and Stykkisholmur/Reykjavik, Iceland.

## 4 Methodology

### 4.1 Processing of MODIS snow products and their validation

The MODIS Re-projection Tool (MRT) (Dwyer et al., 2001) jointly with MODIS Snow Tool (MST) (Gurung et al., 2011) allowed us to:

1. Mosaic the same day tiles h23v05, h24v05, h25v05 for Aqua and Terra separately;

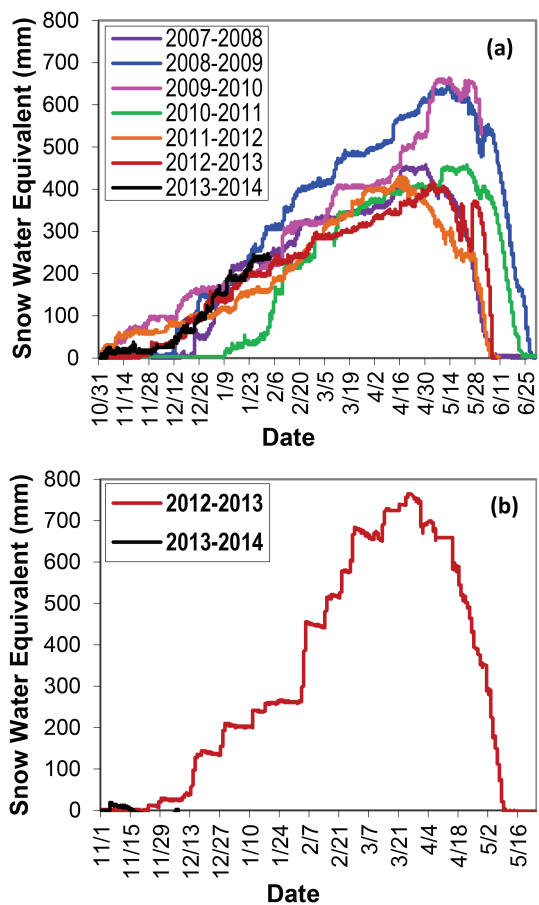

**Figure 3.** Snow water equivalent from snow pillows in the western Himalayas. **(a)** Deosai station – 4149 m a.s.l.; **(b)** Shogran station – 2170 m a.s.l.

2. Re-project twice-a-daily tiles to the Geographical Coordinate System (GCS);

3. Extract area of interest (AOI) covering all study basins.

The errors in the accuracy of MODIS snow products are mainly due to the resemblance between snow and cloud cover. For example, while analysing the MODIS daily snow product (MOD10A1) for the upper Rio Grande River basin for the period 2000–2004, Zhou et al. (2005) found a less than 10 % omission error as a misclassification of snow as land and vice versa. However, they found a high omission error of around 50 % as a misclassification of snow or land as cloud. Therefore, we have validated the considered snow products over the study region prior to their use in our analysis. We have adopted the similar methodology of snow extraction from the Landsat images as applied to the MODIS snow products. We have calculated the NDSI and NDVI from the reflectance of the Landsat TM/ETM+ optical data (bands 2, 3, 4 and 5) and brightness temperatures from the radiance of thermal data such as band 6 for TM and 6-1 for ETM+ (details are in the Supplement). For the non-vegetated areas, we have extracted the snow using a threshold of NDSI > 0.4, while for highly vegetated areas (NDVI > 0.1), a threshold of $0.28 <= \text{NDSI} <= 0.4$ was applied, provided that the reflectance in band 4 was greater than 0.11 and that in band 2 it was greater than 0.10. Mapping of water bodies as snow was avoided using negative NDVI values. Finally, the resultant snow maps were screened to exclude warm objects detected as snow by applying a brightness temperature threshold of greater than 283 K. Snow estimates from the MODIS snow products and the Landsat TM/ETM+ scenes were then compared under clear sky conditions. In order to assess the agreement between the snow estimates of the two data sets, the mean absolute difference (MAD) was calculated using Eq. (1).

$$\text{MAD} = \frac{\sum \left| \text{Snow}_{\text{MYD/MOD}} - \text{Snow}_{\text{Landsat}} \right|}{n} \quad (1)$$

### 4.2 Cloud removal technique and its validation

In order to minimise the cloud data gaps from the validated MODIS snow products, we have applied a rigorous non-spectral cloud removal technique, partially following Wang et al. (2009), Gurung et al. (2011) and López-Burgos et al. (2013). These studies have found the adopted cloud removal technique remarkably efficient in cloud reduction/removal from the MODIS snow products, with few season-dependent trade-offs, resulting in snow maps in good agreement with the ground snow observations. However, these studies suggest numerous cloud removal steps, where each successive step removes more cloud but is subject to a high probability of information loss (Gafurov and Bárdossy, 2009). We select a few but robust cloud removal steps and briefly discuss here their functionality and theoretical accuracy.

In the first step, we have merged both MODIS Terra and Aqua same day images in order to remove short-persisting clouds. We take the MODIS Terra snow product as base as it experiences relatively less cloud cover than its counterpart (Parajka and Blöschl, 2008; Wang et al., 2009). We replace the cloudy pixels of the MODIS Terra snow images with the corresponding cloud-free pixels of same day Aqua images. This step results in a single snow image per day. This step is most effective and accurate as both the observations are only few hours apart – just 3–4 h at the Equator – (Gafurov and Bárdossy, 2009).

As for temporal filling, we replace the present day cloudy pixel with a corresponding previous day snow pixel. In temporal analysis, if both the previous and next day corresponding cloud-free pixels observe the same class, the present day corresponding cloudy pixel is replaced by that class. Here, we assume that the snow cover often persists more than one day, thus the present-day cloudy pixel can be replaced with the corresponding previous day snow pixel. If the previous-day snow observation was taken from the Aqua image as a result of the first step of the cloud removal technique then it is only few hours apart from the present day cloudy pixel. During such time, three possibilities of either rapid snowmelt, snow fall or no change can occur. Possibility of rapid snowmelt is limited because the present day is cloudy, blocking incoming solar radiation (Gafurov and Bárdossy, 2009), while snow fall or no change possibilities still favour our hypothesis. Though the MODIS Terra and Aqua snow products are in good agreement, these products still have little differences, particularly during the transitional period (Wang et al., 2009). These differences are attributed to change of a pixel from land to snow in case of snow fall and snow to land in case of rapid melting during the time between MODIS Terra and Aqua acquisitions (Prajka and Blöschl, 2008). Snow redistribution by a wind blow can also result in differences among Terra and Aqua acquisitions. Effects of such sub-daily variations cannot be accounted for in this step due to the limited availability of sub-daily scale observations, and because our analysis considers only one observation per day. This information loss even exists under clear sky conditions.

As for spatial filling, we have replaced a particular cloudy pixel with the dominant class observed among its neighbouring pixels. Such a replacement cannot always be accurate, but its probability of being accurate is high. A major problem can be the patchiness of snow, resulting in spurious information. Therefore, we have considered the spatial window size of $7 \times 7$, in order to decide the new class of the cloudy pixel based on a reasonable number of neighbouring pixels. Gurung et al. (2011) have reported this window size as optimal for the spatial filling in order to remove/reduce cloud cover from the MODIS snow products. We first combined MODIS Terra and Aqua images, and then applied temporal filling, spatial filling, and temporal analysis. In order to avoid the influence of the remaining cloud cover, daily snow cover estimates for each basin with less than 10 % cloud cover only were considered for further analysis.

We have validated our adopted cloud removal technique before its use to improve the snow products. Since we only have SWE observations (2008–2012) from the Deosai station, we have decided to perform both absolute and relative validation. For absolute validation, we compare the MODIS snow products against the Deosai SWE observations (Fig. 3). The installed snow pillow measures incident snow in units of 3.18 mm water equivalent. Therefore, we have chosen a threshold of SWE $>=3.18$ mm to decide the existence of snow, and a threshold of SWE $=0$ for clear land/no snow. We have provided confusion matrices for the absolute validation (Table 2), as well as the estimates of overall accuracy (Eq. 2) and snow miss (Eq. 3) and false alarm (Eq. 4) rates.

$$\text{Overall accuracy} = \frac{(a_1 + b_2)}{(a_1 + b_1 + a_2 + b_2)} \quad (2)$$

**Table 2.** Confusion matrices for the MODIS snow products against in situ SWE observations at the Deosai station for the period 2008–2012.

| Observations<br>MODIS snow product(s)↓ | Snow | No snow/<br>clear land |
|---|---|---|
| Snow (SWE $>=3.18$ mm) | a1 | b1 |
| No snow/clear land (SWE $=0$ mm) | a2 | b2 |

$$\text{Snow miss rate} = \frac{(b_1)}{(a_1 + b_1 + a_2 + b_2)} \quad (3)$$

$$\text{False alarm rate} = \frac{(a_2)}{(a_1 + b_1 + a_2 + b_2)} \quad (4)$$

$a_1$ represents the total number of correct snow hits, and $b_2$ represents the total number of correct land/no-snow hits. The variable $b_1$ represents the total number of occurrences when the MODIS product indicates no snow/clear land, but ground observation shows the existence of snow. The total number of situations where the MODIS product indicates the existence of snow, but the observation suggests no snow/clear land, is represented by the variable $a_2$.

For the relative validation, the year 2004 was taken as a validation period for a two-fold reason; it was the first wet year after a long drought (1998/99–2002/03) over the Indus basin (Levinson and Waple, 2004; Baig and Rasul, 2009), and it experienced the maximum cloud cover conditions. Within the validation period, we have taken ten pairs of same-day Terra and Aqua snow images that feature large cloud cover differences, and become nearly cloud free when the cloud removal procedure is applied. Note that the same-day Terra and Aqua images are two independent observations, acquired at different times of the day but processed with a similar approach (Parajka and Blöschl, 2008). We first make the Terra snow images cloud free by masking out their cloud covers, and then apply these Terra cloud masks to the corresponding same-day Aqua images. At this stage, the rest of the cloud cover of the Aqua snow images represents the areas which are cloud free in the corresponding same-day Terra snow images. We apply all the steps of the cloud removal technique, except the first one, to the new Aqua snow images to remove their remaining cloud cover. The resultant snow and cloud cover estimates are then compared to the estimates from the same-day Terra snow images, and MAD was calculated using Eq. (1). We considered the performance of our cloud removal technique to be satisfactory if it reduced the cloud cover to less than 10 % of AOI for the validation period.

### 4.3 Snow cover estimates and precipitation distribution

We have estimated the snow cover against elevation, slope and aspect from the improved snow product by dividing the basin areas into 500 m elevation zones, 10-degree slope zones and 22.5-degree aspect zones (Figs. 11 and 13). We have ascertained the mean seasonal snow cover trends for the four time slices, namely the autumn (September–November), winter (December–February), spring (March–May) and summer (June–August) seasons, in addition to mean annual trends for all the studied basins. It is pertinent to mention here that the unavailability of SD/SWE from the observations and the unsuitability of such parameters derived from remotely sensed data sets have restricted our analysis to snow cover assessment only. However, we have analysed precipitation records from 36 low- and high-altitude stations on seasonal and annual timescales in order to present a comprehensive picture of moisture input to the main parts of the study region. We also estimate the winter/spring SWE by accumulating precipitation from the SIHP stations for the period during which temperatures contiguously remained below zero.

In order to obtain a qualitative mass balance response of the existing glaciers to the prevailing climatic regime of the study area, we estimate the end-of-summer regional SLA zones from height-dependent snow cover estimates, and ascertain their tendencies. Alternatively, variations in the accumulation area ratio (AAR), in general, can also be used in such regards. However, a number of glaciers in our high-relief study area feature the avalanche-fed accumulation regime (Hewitt, 2011). For such glaciers, AAR cannot always be related successfully to their mass balances (e.g. the Dunagiri and Changme–Khangpu glaciers in the Himalayas), especially when a short time series of AAR is analysed (Dyurgerov et al., 2009). The balanced-budget AAR for these glaciers, however, is reported to be less than 0.5 (Kulkarni, 1992; Braithwaite and Raper, 2009). In view of such limitations of AAR for the study area, we have analysed the end-of-summer regional SLA zone. We also compare the inter-annual variability of the SLA zone against the median elevation of basin-hosted glaciers, which is a reasonable topographical data-based proxy for the long-term equilibrium line altitude (ELA – the altitude where net mass gain/loss is zero) of the glaciers (Braithwaite and Raper, 2009).

### 4.4 Tele-connections

The hydrology of the studied basins dominates, with moisture inputs from westerly disturbances. The frequency, strength and track of the westerly disturbances are influenced by the pressure conditions over the North Atlantic, which can be explained by the North Atlantic Oscillation (NAO) index (Hurrell, 1995; Syed et al., 2006). The variation in the NAO index affects precipitation and temperature over northern Europe (Fowler and Kilsby, 2002) and – at the furthermost extent – over the study region. Syed et al. (2006) explain that during the positive NAO phase, the southern flank of the Mediterranean storm track becomes intensified over northern Iran and Afghanistan due to the enhanced low pressure over Afghanistan and central Asia. Such conditions possibly feature an additional moisture transport from the Caspian and Arabian seas – resulting in a positive precipitation anomaly – to the study area as well as to a large part of the central southwestern Asia region. The El Niño–Southern Oscillation (ENSO) during its warm phase also leads to similar conditions over the study region (Shaman and Tziperman, 2005; Syed et al., 2006). Here, considering the atmospheric-only mode of the global index, i.e. NAO, we explore whether such an index can partly explain the variability of snow cover with a possible lead time and whether such variability contributes further to the variability of the melt-season stream flow. Our analysis is based on studying the simple Pearson correlation between two time series of equal length (2001–2012). We have correlated the seasonal NAO index for several seasons with the winter season snow cover, in order to explore the effect of changing the lead time.

## 5 Results

### 5.1 Validation of MODIS snow products

Our validation results show that MAD between the snow cover estimates from MODIS and Landsat is around 2 % of the mean investigated area for both Terra and Aqua products (Fig. 4, Table 3). Whereas, the relative difference (difference in MODIS and Landsat snow cover divided by Landsat snow cover) for each individual pair ranges from 0.1 to 25 % (Table 3). We have found that MODIS slightly overestimates the snow cover during the spring and summer seasons, while it underestimates it during the autumn season, relative to the Landsat images. Tang et al. (2013) have also found such an overestimation from the MODIS product for the Mount Everest region. The underestimation, however, has been observed under highly patchy snow conditions, mainly over the Jhelum basin (Path 149, Row 36), and during the autumn season (days of year 276 and 290). In fact, the fine resolution of Landsat TM/ETM+ as compared to MODIS has allowed the precise detection of such patchy snow cover. Overall, we have found the MODIS snow cover estimates in good agreement with the Landsat data under clear sky conditions.

### 5.2 Validation of cloud removal technique

Our absolute validation shows that under clear sky conditions, the MODIS Terra and Aqua snow products feature an overall accuracy of 86 % and around 85 %, respectively, against the observations at Deosai station. Both the snow miss and false alarm rates for the MODIS Terra product are around 7 %. For the MODIS Aqua product, false alarm rate is around 6 % while the rate of snow miss is around 9 %. Relatively higher snow miss rate for the MODIS Aqua product is due to its larger cloud coverage observed as compared to the MODIS Terra product – 40 % against 36 % at Deosai location for the period 2008–2012.

Table 3. Results of the MODIS Terra and Aqua product validation against the Landsat images under clear sky conditions.

| S. no. | Landsat scene identifier | MODIS Terra snow (km$^2$) | Landsat snow (km$^2$) | Relative diff. (%) | Absolute diff. (%) | MODIS Aqua snow (km$^2$) | Landsat snow (km$^2$) | Relative diff. (%) | Absolute diff. (%) |
|---|---|---|---|---|---|---|---|---|---|
| 1 | LE71490362001097SGS00 | 15 117 | 15 535 | −3 | 1.3 | – | – | – | – |
| 2 | LE71500362001120SGS00 | 4592 | 4244 | 8 | 1.1 | – | – | – | – |
| 3 | LE71480352001138SGS00 | 13 120 | 12 909 | 2 | 0.6 | – | – | – | – |
| 4 | LE71510352001143SGS00 | 8107 | 7592 | 7 | 1.6 | – | – | – | – |
| 5 | LE71490352002100SGS00 | 20 365 | 19 225 | 6 | 3.5 | – | – | – | – |
| 6 | LE71490362002100SGS00 | 19 344 | 19 309 | 0 | 0.1 | – | – | – | – |
| 7 | LE71470352002102SGS00 | 7225 | 7755 | −7 | 1.6 | – | – | – | – |
| 8 | LE71500352002155SGS00 | 10 116 | 9931 | 2 | 0.6 | – | – | – | – |
| 9 | LE71470362002214SGS00 | 4174 | 3468 | 20 | 2.1 | – | – | – | – |
| 10 | LE71490362002276SGS00 | 1269 | 1693 | −25 | 1.3 | 944 | 1073 | −12 | 0.4 |
| 11 | LE71490352003119EDC00 | 21 101 | 19 653 | 7 | 4.4 | 21 410 | 19 681 | 9 | 5.2 |
| 12 | LT51500362009262KHC00 | 782 | 830 | −6 | 0.1 | 355 | 827 | −57 | 1.5 |
| 13 | LT51490362010146KHC00 | 12 582 | 10 431 | 21 | 6.6 | 10 845 | 10 365 | 5 | 1.5 |
| 14 | LT51490362010290KHC00 | 1174 | 1506 | −22 | 1.0 | 810 | 1490 | −46 | 2.1 |
|  |  |  |  | MAD Terra | 1.85 |  |  | MAD Aqua | 2.14 |

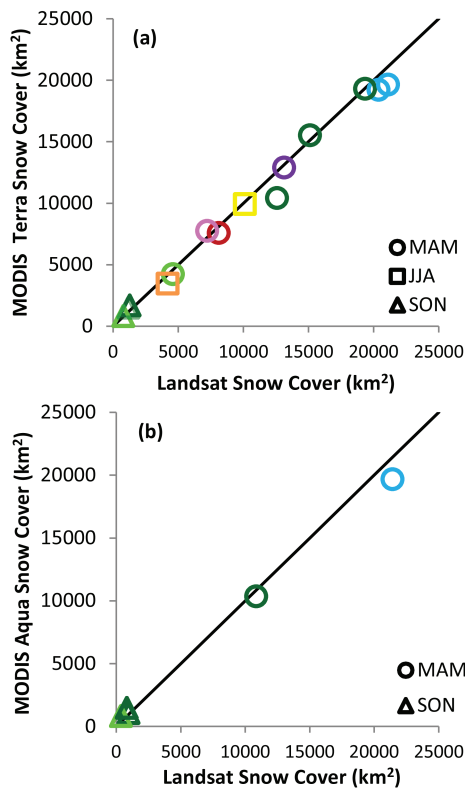

Figure 4. Validation results of the MODIS snow products against Landsat TM/ETM+ images. (a) MODIS Terra product against Landsat scenes; (b) MODIS Aqua product against Landsat scenes. Note: each colour identifies a unique location ("Path" and "Row") of the Landsat scene acquisition, while each different shape identifies a particular season.

Our cloud removal procedure fills up 77 % of the cloud data gaps in the MODIS Terra product, and around 79 % of the cloud data gaps in the MODIS Aqua product, with an accuracy of 90 % against the observations. The procedure also improves the overall accuracy of the MODIS Terra (86 %) and Aqua (85 %) products to 88 %. In individual steps, combining the MODIS Terra and Aqua products fills up 16 % of the total filled data gaps with an accuracy of around 89 % against the observations (Table 4). The false alarm and snow miss rates for this step are around 4 and 7 %, respectively. The temporal filling, spatial filling and temporal analysis steps have filled up around 65, 8 and 14 % of the total filled data gaps, respectively, with an accuracy of 90 % against the observations. Their false alarm rate is around 4 %, while the snow miss rate is around 6 %. The snow miss and false alarm rates for the filtered MODIS product are mostly observed during the months of June and November. We have observed that the snow miss rate is highly sensitive to the applied SWE threshold in order to decide the presence of snow – if we increase the threshold, the snow miss rate drops considerably.

Table 5a and b shows that clouds from the Aqua images have been reduced considerably, and the snow estimates have consequently been improved, which are now comparable to the MODIS Terra snow estimates. Figure 5 spatially illustrates the validation of the applied technique for day 96 of the year 2004. The MAD between the MODIS Terra and Aqua snow estimates is around 0.54 %. This may partly be attributed to the spatial cloud/snow cover differences between the same-day Terra and Aqua images (as depicted by Table 5b). It is also partly due to the fact that cloud has not been completely removed from the Aqua images, because the first step in combining both same-day images has been skipped during this validation process. During the whole validation period of the year 2004, the applied cloud removal technique has reduced the cloud cover from 37 % (MOD10A1) and 43 % (MYD10A1) to 7 %, improving snow cover estimates from 7 % (MOD10A1) and 5 % (MYD10A1) to 14 % for the whole AOI. In the individual steps, combining the MODIS Terra and Aqua same-day images, the temporal filling, the spatial filling and the temporal analysis have reduced the average cloud cover to 29, 9, 8 and 7 %, and have improved the snow cover to around 8, 12, 13 and 14 % of the whole AOI, respectively (Figs. 6a, b and 7).

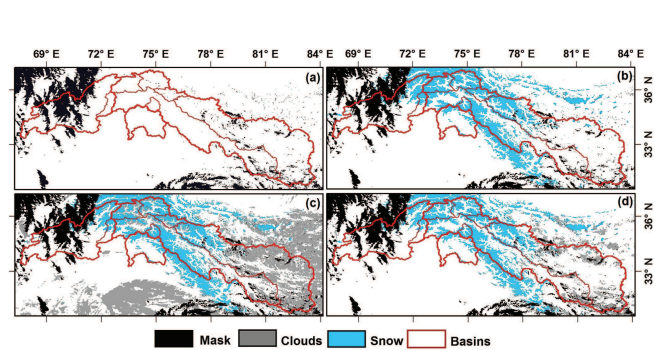

**Figure 5.** Spatial illustration of the cloud removal technique validation over the Aqua snow image with respect to the Terra snow image for day 96 of the year 2004 over the whole AOI (Terra cloud mask in black, snow cover in blue, and the Aqua cloud cover is shown in grey). **(a)** Terra cloud cover, **(b)** Terra snow image after masking out the Terra cloud cover, **(c)** Aqua snow image after masking out the Terra cloud cover, **(d)** Aqua snow image after applying all cloud removal steps, except for combining it with the Terra same-day image.

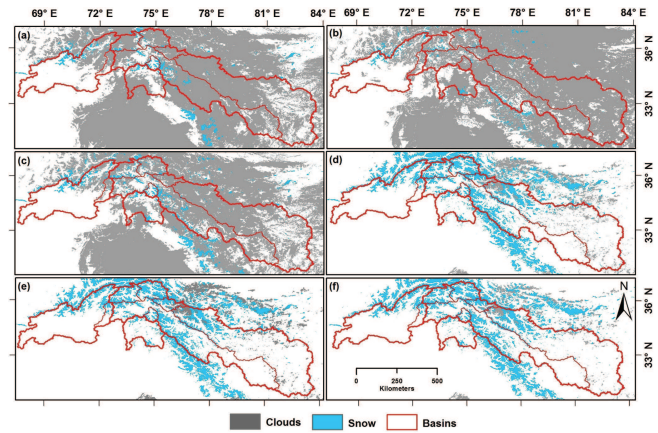

**Figure 7.** Cloud and snow cover of original MODIS Terra and Aqua Images and after implementation of five steps for day 144 of year 2004, **(a)** original MODIS Terra, **(b)** original MODIS Aqua, **(c)** after combining Terra and Aqua, **(d)** after temporal filling, **(e)** after spatial filling, and **(f)** after temporal analysis.

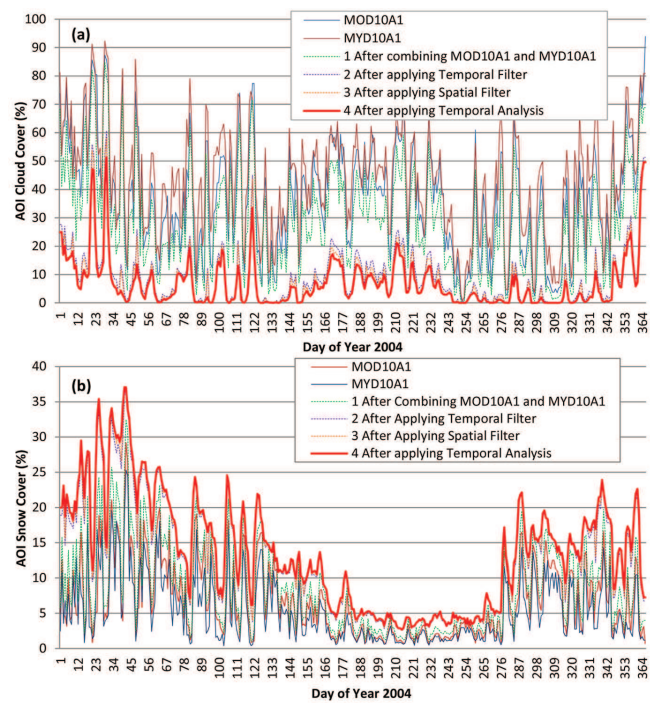

**Figure 6.** **(a)** Cloud coverage from the MODIS Terra (MOD10A1), Aqua (MYD10A1) and after each step of the cloud removal Technique for the validation period of year 2004. MOD10A1 and MYD10A1 are original cloud cover estimates from MODIS Terra and Aqua, respectively, while steps 1–4 indicate the cloud cover estimates after combining both Terra and Aqua products, and after applying the temporal filling, spatial filling and temporal analysis, respectively. **(b)** Same as Fig. 6a, but for snow cover estimates.

### 5.3 Basin-wide snow cover estimates and precipitation distribution

Note that our analysis was based on the snow cover estimates for the days experiencing cloud cover less than 10 % of the respective basin areas. Such days accounted for 65–85 % of the total number of days over the whole period for all basins, except Shigar and Hunza sub-basins. We have observed that winter (30–65 %) and spring (60–80 %) seasons have the minimum number of days with less than 10 % cloud cover, whereas for summer and autumn seasons, the number of such days was always higher (i.e. 60–90 %). The Shigar and Hunza (high-latitude/altitude glacierised) sub-basins have experienced only 45–50 % of days of less than 10 % cloud cover for the whole period, whereas such days during the winter season are about 15 % for the Shigar sub-basin and about 25 % for the Hunza sub-basin.

#### 5.3.1 Precipitation distribution

In Fig. 3, we show snow-pillow-based measurements of SWE from the Deosai and Shogran stations in the western Himalayas. At the Deosai station, SWE ranges between 400 and 700 mm during the period 2008–2013, while for the

Table 4. Results of the absolute validation of the cloud removal technique; MODnMYD refers to MOD10A1 and MYD10A1 combined images, TF refers to temporally filled images, SF refers to spatially filled images, and TA refers to temporally analysed images.

| Clear sky conditions | Cloudy days | Snow miss rate (%) | False alarm rate (%) | Overall accuracy against observations (%) |
|---|---|---|---|---|
| MODIS | 895 | 7.2 | 6.8 | 86.0 |
| MYD | 946 | 9.0 | 6.2 | 84.8 |
| MODnMYD | 750 | 6.9 | 6.5 | 86.7 |
| TF | 258 | 6.3 | 5.6 | 88.1 |
| SF | 238 | 6.3 | 5.6 | 88.0 |
| TA | 203 | 6.3 | 5.4 | 88.2 |
| MODIS cloudy conditions | Filled cloudy days | Snow miss rate (%) | False alarm rate (%) | Accuracy of filled data by cloud removal technique against observations (%) |
| MODnMYD | 16 | 7.1 | 4.3 | 88.6 |
| TF | 65 | 5.9 | 3.9 | 90.2 |
| SF | 8 | 5.8 | 4.0 | 90.1 |
| TA | 14 | 6.0 | 3.7 | 90.3 |

Table 5. (a) Relative validation: snow cover estimates (%) for AOI, actual and after applying each cloud removal step for the selected dates within the validation period of the year 2004. Note: MAD is 0.54. (b) Same as (a), but for the cloud cover estimates.

(a)

| S. no. | DOY (2004) | Actual | | MYD snow after | | | | MYD and MOD snow difference |
|---|---|---|---|---|---|---|---|---|
| | | MOD snow | MYD snow | masking | TF | SF | TA | |
| 1 | 85 | 21.3 | 8.7 | 9.3 | 19.9 | 20.0 | 20.1 | 1.2 |
| 2 | 96 | 12.5 | 9.2 | 9.8 | 12.7 | 12.8 | 12.8 | 0.3 |
| 3 | 97 | 15.5 | 7.9 | 10.4 | 15.1 | 15.2 | 15.2 | 0.2 |
| 4 | 124 | 18.4 | 10.8 | 12.3 | 17.8 | 18.0 | 18.1 | 0.3 |
| 5 | 129 | 7.3 | 2.4 | 3.4 | 7.7 | 7.7 | 7.7 | 0.4 |
| 6 | 212 | 2.4 | 0.8 | 1.0 | 2.0 | 2.0 | 2.1 | 0.3 |
| 7 | 271 | 4.6 | 2.8 | 3.0 | 3.9 | 3.9 | 4.0 | 0.7 |
| 8 | 282 | 6.4 | 3.3 | 4.0 | 6.2 | 6.2 | 6.2 | 0.2 |
| 9 | 290 | 13.3 | 7.3 | 7.1 | 12.2 | 12.2 | 12.3 | 1.0 |
| 10 | 298 | 13.8 | 9.3 | 9.7 | 14.5 | 14.5 | 14.5 | 0.7 |

(b)

| S. no. | DOY (2004) | Actual | | MYD cloud after | | | | MYD and MOD cloud difference |
|---|---|---|---|---|---|---|---|---|
| | | MOD cloud | MYD cloud | masking | TF | SF | TA | |
| 1 | 85 | 15.3 | 28.9 | 20.2 | 0.7 | 0.4 | 0.3 | 0.25 |
| 2 | 96 | 12.2 | 30.5 | 26.5 | 5.7 | 4.8 | 4.8 | 4.81 |
| 3 | 97 | 31.7 | 52.0 | 36.1 | 0.3 | 0.2 | 0.1 | 0.07 |
| 4 | 124 | 17.7 | 31.0 | 21.7 | 1.9 | 1.5 | 0.7 | 0.74 |
| 5 | 129 | 44.2 | 54.7 | 34.8 | 0.5 | 0.4 | 0.2 | 0.24 |
| 6 | 212 | 58.2 | 59.3 | 37.3 | 9.0 | 7.1 | 1.0 | 1.03 |
| 7 | 271 | 16.9 | 29.7 | 20.3 | 0.3 | 0.2 | 0.1 | 0.09 |
| 8 | 282 | 25.3 | 38.4 | 21.8 | 0.1 | 0.1 | 0.0 | 0.03 |
| 9 | 290 | 13.5 | 23.5 | 16.4 | 0.6 | 0.4 | 0.1 | 0.07 |
| 10 | 298 | 8.8 | 26.0 | 20.8 | 0.2 | 0.1 | 0.1 | 0.07 |

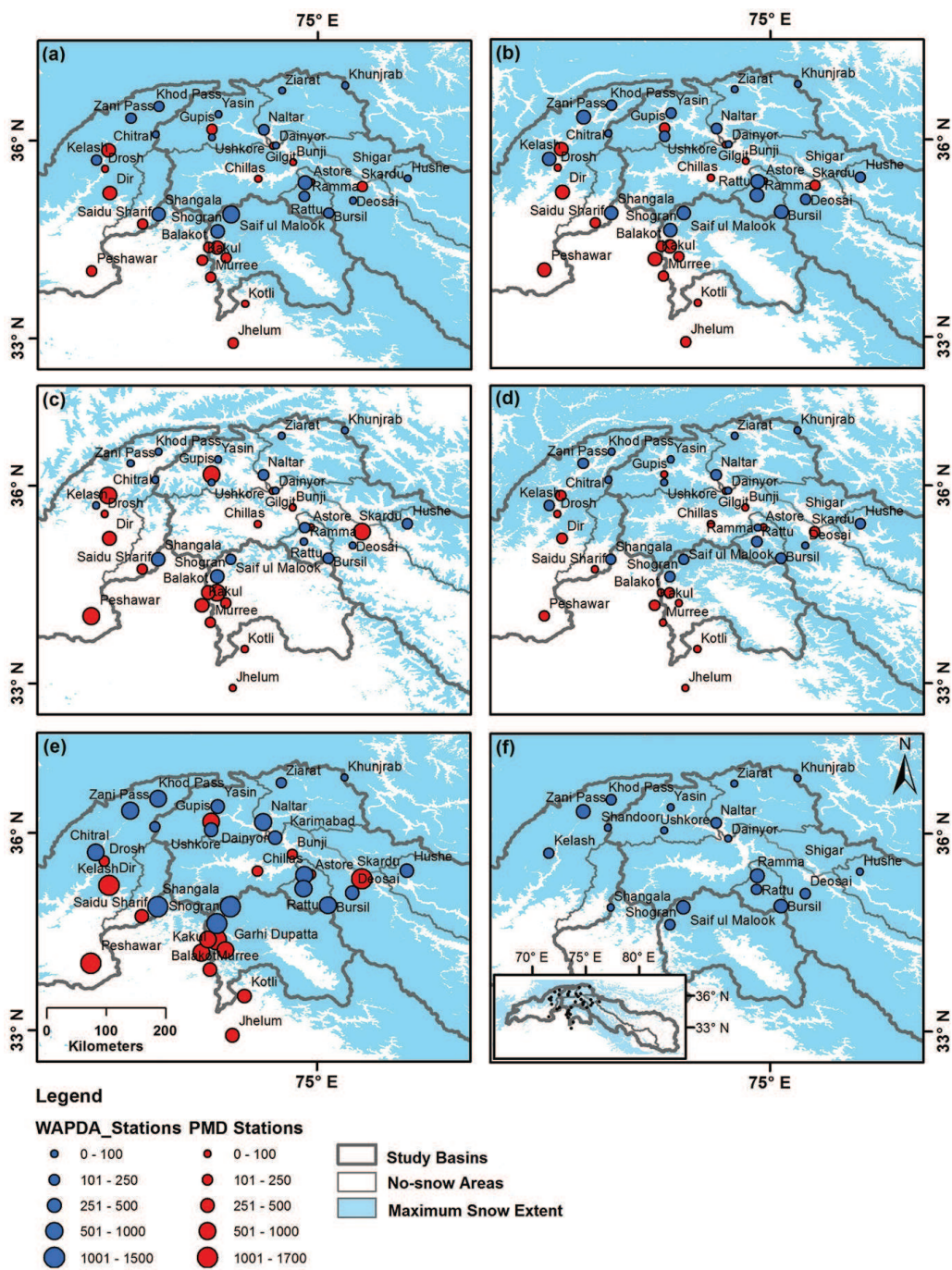

**Figure 8.** Seasonal and annual precipitation climatology from low-altitude (below 2200 m a.s.l.) PMD stations (1995–2010) and from high-altitude stations (above 2200 m a.s.l.) WAPDA stations (1995–2012) for: **(a)** winter, **(b)** spring, **(c)** summer, **(d)** autumn, **(e)** annual, **(f)** climatology of annual precipitation accumulated during the period when temperatures contiguously remained below zero.

Shogran station, the SWE is above 700 mm for the years 2012–2013. Figure 8a–f shows that the region receives most of its moisture during the winter and spring seasons, while there is relatively little moisture input during the autumn season. The ratio of winter and spring precipitation to the total annual precipitation ranges from 43 to 76 % for high-altitude (above 2200 m a.s.l.) stations (WAPDA), and from 26 to 77 % for low-altitude (below 2200 m a.s.l.) stations (PMD). In the summer season, moisture input to the region (the Jhelum, Kabul, Swat and Gilgit basins/sub-basins) is dominated by the south Asian summer monsoonal precipitation regime. The annual average precipitation ranges from below 50 to around 1700 mm for the low-altitude stations, and from around 200 to around 1400 mm for high-altitude stations (Fig. 8e). Within the Karakoram range, we have found an annual average precipitation range from 200 to 700 mm at the Khunjrab and Naltar stations, respectively. For the western Himalayas, this range is from 150 to 1400 mm approximately at the Astore and Saif ul Maluk stations, respectively. Within the Hindu Kush range, annual average precipitation ranges from below 50 to around 1700 mm at the Gilgit and Chitral stations, respectively.

Table 6. List of nine basins with total areas along with snow coverage (SC) % and trend slopes (2001–2012). Values in bold are statistically significant; values in italic show a variability greater than or equal to 5 %.

| S. no. | Basin at gauging site | Snow cover % | | | Snow cover % trend slope | | | | |
|---|---|---|---|---|---|---|---|---|---|
| | | Mean Min ± SD | Mean Max ± SD | Avg. ± SD | DJF | MAM | JJA | SON | Ann. |
| 1 | Astore at Doyian | 2 ± 1 | 98 ± 1 | *47 ± 5* | −0.79 | +0.49 | +0.76 | −0.72 | −0.29 |
| 2 | Gilgit at Gilgit | 3 ± 1 | 90 ± 4 | 41 ± 4 | −0.58 | +0.50 | +0.77 | +0.02 | +0.16 |
| 3 | Hunza at Dainyor Bridge | *17 ± 6* | 83 ± 4 | 49 ± 3 | −1.07 | +0.09 | +0.38 | −0.30 | −0.12 |
| 4 | Jhelum at Azad Patan | 0.2 ± 0.2 | *77 ± 8* | 22 ± 2 | +0.16 | +0.47 | **+0.33** | +0.23 | +0.30 |
| 5 | Kabul at Nowshera | 1 ± 0.3 | 67 ± 5 | 18 ± 2 | −0.05 | +0.11 | +0.19 | −0.01 | +0.12 |
| 6 | Swat at Chakdara | 1 ± 0.5 | *72 ± 9* | 27 ± 3 | −0.04 | +0.39 | +0.37 | −0.09 | +0.15 |
| 7 | Shigar at Shigar | *25 ± 8* | 90 ± 3 | 58 ± 3 | −0.76 | +0.30 | +0.37 | −0.17 | −0.02 |
| 8 | Shyok at Yugo | 3 ± 1 | *44 ± 9* | 14 ± 2 | −0.21 | −0.63 | +0.09 | −0.16 | −0.17 |
| 9 | UIB at Besham Qila | 4 ± 1 | *54 ± 7* | 21 ± 2 | −0.74 | −0.07 | +0.21 | −0.13 | −0.15 |

Our estimated winter/spring SWE (Fig. 8f) suggests that the maximum solid moisture input to the region is around 400 mm at Bursil and Saiful Maluk stations. For most of the high altitude stations, estimated SWE is either similar or higher than their total winter precipitation. The case of higher accumulated precipitation indicates that the snow accumulation at these locations extends well into the spring season. We have noted that few stations at lower latitudes, such as, Shangala, Shogran and Saif ul Maluk observe the opposite case, indicating smaller amount of incident snow fall (Fig. 8f) at these locations.

### 5.3.2 Snow cover

Shigar has the highest annual average snow cover percentage, followed by Hunza, Astore and Gilgit (Fig. 9, Table 6). These sub-basins have shown large year-to-year variation as compared to other lower-latitude/altitude basins. On the other hand, the Shyok sub-basin has the lowest snow cover percentage, mainly due to its large extent at lower latitudes in the southeast.

The snow cover for the Astore and Gilgit sub-basins of the UIB ranges from 2 ± 1 to 3 ± 1 % during the summer season, and from 98 ± 1 to 90 ± 4 % during the spring season, respectively. These basins experience relatively low cloud cover, but high variability in both the accumulation and ablation seasons. The sharp drop in the snow depletion curve implies that both of the basins have a low glacier melt contribution, so that their hydrology mainly depends upon the snowmelt (Fig. 10a–b). This is further clarified by the minute minimum snow cover (near 0 %) and the small glacier cover reported for these basins.

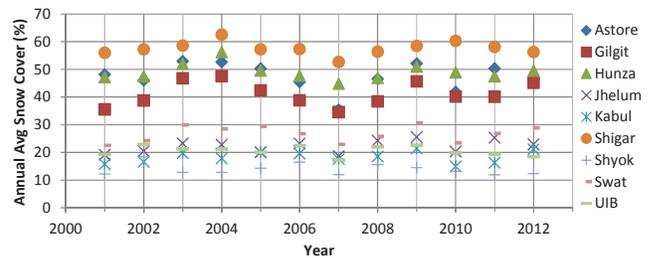

Figure 9. Annual average snow cover for all basins for the period 2001–2012.

For the Hunza and Shigar stations, the snow cover extent ranges from 17 ± 6 to 25 ± 8 % as the minimum during summer, and from 83 ± 4 to 90 ± 3 % as the maximum during spring, respectively. We can observe a relatively smooth drop of the snow depletion curve during the ablation period (Fig. 10c–d). Conversely to what we have observed in the Astore and Gilgit sub-basins, here the sub-basins feature relatively low snow cover variability throughout the year, which may be associated with their particular topographic characteristics. These basins extend into relatively high-latitude/altitude zones that feature the hydro-climatic conditions resulting in perennial snow and permanent ice cover.

The snow cover for the Jhelum and Kabul basins ranges from about 1 % for both basins during summer to 75 ± 8 and 67 ± 5 % during the spring season, respectively. These two basins exhibit high snow cover variability during the accumulation period and low snow cover variability during the ablation period (Fig. 10e–f). We have observed great similarity among the Jhelum and Kabul River basins in terms of their minimum and maximum snow coverage and snow depletion patterns (Fig. 10e–f). This observation is further reinforced by the fact that both river basins provide a similar contribution (about 16 %) to the annual average surface water available in Pakistan (Ali et al., 2009). Swat, a comparatively smaller sub-basin of the Kabul basin, receives snow coverage ranging from $1 \pm 1$ to $72 \pm 9$ %. Its snow accumulation/ablation patterns and other snow cover characteristics are similar to those of the Kabul and Jhelum basins (Fig. 10i).

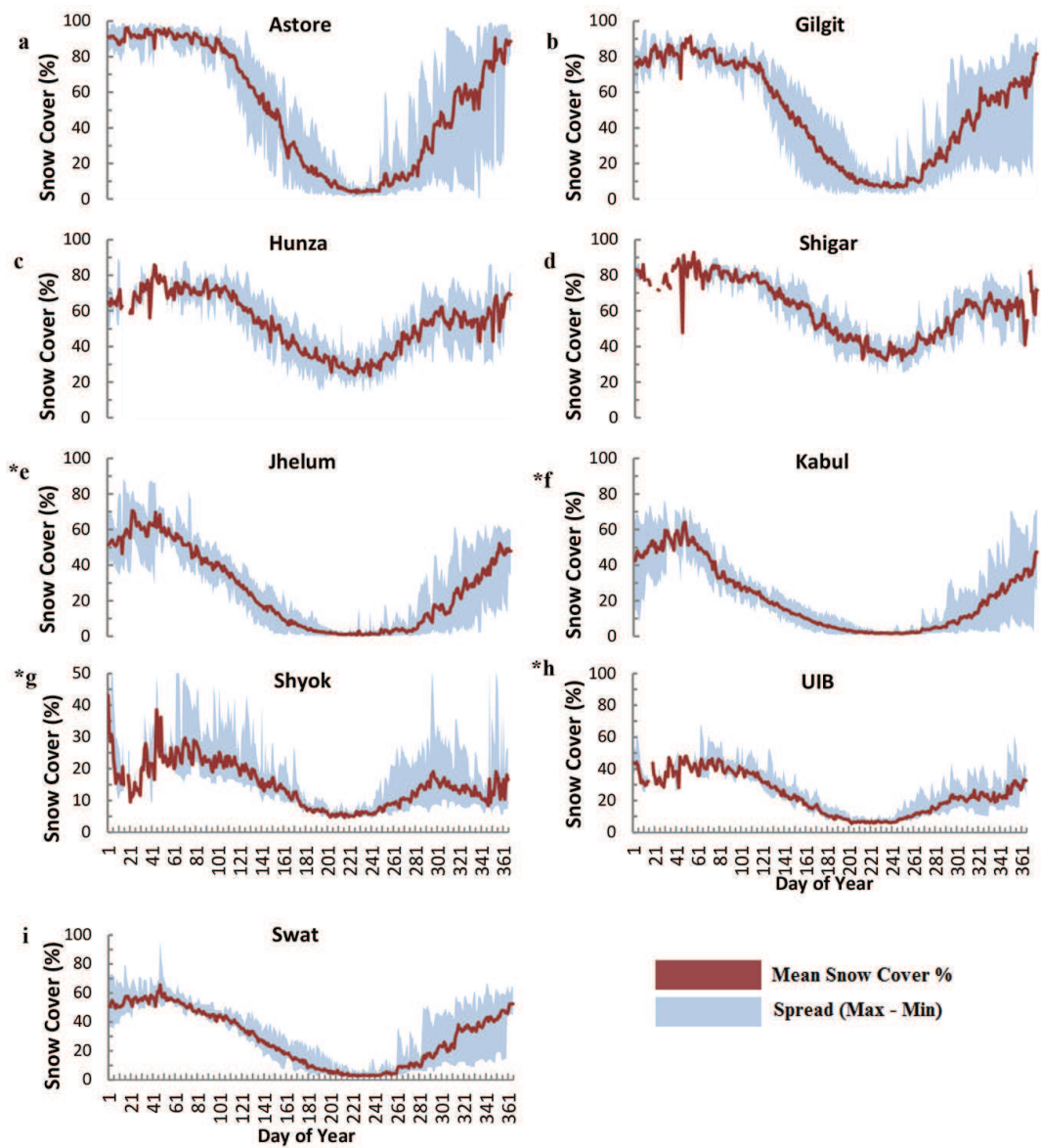

**Figure 10. (a–i)**: Mean snow cover and its variability over the period 2001–2012 for all basins. The blue shaded area shows the spread (mean minimum and mean maximum), whereas the red line shows the mean of the snow cover percentage over the whole period. Note: (*) indicate major basins.

The mean annual cycle of the snow cover percentage for the UIB and its major sub-basin Shyok has been found to be quite different from that of the other basins. The snow cover of these basins ranges from $4 \pm 1$ to $3 \pm 1$ % in the summer, and from $54 \pm 7$ to $44 \pm 9$ % in the spring season, respectively (Fig. 10g–h). Both of the basins showed high snow cover variability in accumulation and ablation seasons. Such variability was substantially pronounced for the Shyok sub-basin throughout the year except during winter season that feature large and persistent cloud cover. A large portion of the northeastern Shyok sub-basin, characterised by a low precipitation rate, has neither permanent snow nor a glacier cover. Instead, the part of the Shyok sub-basin lying in the Karakoram range features a high concentration of ice (Hewitt, 2007; Bhambri et al., 2013). Hence, its contribution to the stream flow mainly comes from the glacier melt. For the UIB, the minimum snow cover corresponds to the estimates of its sub-basins Gilgit, Hunza, Shigar, Shyok and UIB itself, so it exhibits the average effect of the contrasting hydrological regimes of its sub-basins.

The analysis of annual data depicted consistency with the worst drought in Pakistan, which spanned from 1998/99 to 2002/03 and weakened in 2003–2004 due to heavy winter precipitations (Levinson and Waple, 2004; Baig and Rasul, 2009). Interestingly, our extracted mean annual minimum snow cover extent for all basins except Astore and Hunza generally corresponds to the areal extent of the existing glaciers, with a slight systematic underestimation (Table 1).

Most of the mid-altitude basins, namely Jhelum, Kabul and UIB including Swat and Shyok sub-basins show a variability $\geq 5$ % in their mean maximum snow cover. The Astore and Gilgit sub-basins show high variability in their mean snow cover. Among high altitude basins, only Shigar sub-basin shows high variability in its mean minimum snow cover. A large spread is found during the snow accumulation and ablation seasons. Such variability is typically higher for the snow-fed basins than for the glacier-fed basins. High variability in the minimum, maximum and the annual average snow coverage directly affects the melt-water runoff contribution on yearly basis, which may further contribute to the inter-annual variability of the IRS flows.

### 5.3.3 Snow cover trends

For the annual average snow cover, we have found a slightly decreasing trend for UIB and for all the sub-basins, except Gilgit, and a slightly increasing trend for the rest of the studied basins, though no trend was statistically significant (Table 6). The hydrology of the basins showing a decreasing trend is mainly influenced by the westerly disturbances, while for the basins showing an increasing trend, it is mainly influenced by the south Asian summer monsoon. On a seasonal timescale, the winter and autumn seasons feature a decreasing trend for most of the studied basins. Instead, there was an increasing trend for the Gilgit basin in the autumn season and for the Jhelum basin throughout the year. Most basins show an increasing trend in the rest of seasons, except for Shyok and UIB, which have shown a decreasing trend for the spring season. However, a statistically significant trend was found only for the Jhelum basin in the summer season.

### 5.3.4 Height dependence of snow cover estimates

We find high seasonal variation of the height-dependent snow cover for the snow-fed basins as compared to the glacier-fed basins throughout the year (Fig. 11). The maximum snow cover for the high-altitude/latitude basins is observed during winter, while for lower latitude/mid-altitude basins, it occurs during the spring season (Fig. 11). In most of the basins, a disproportionately large fraction of the snow cover comes from high-altitude zones, which include very small surface areas, such as zones above 5000 m a.s.l. for the Astore, Swat and Jhelum basins, above 5500 m a.s.l. for the Kabul basin, above 6000 m a.s.l. for the Gilgit, Shyok and UIB basins, and above 6500 m a.s.l. for the Hunza and Shigar basins. Conversely, the Jhelum, Kabul and Swat basins have considerable surface areas below 2000 m a.s.l., which have a negligible snow cover percentage (Fig. 11).

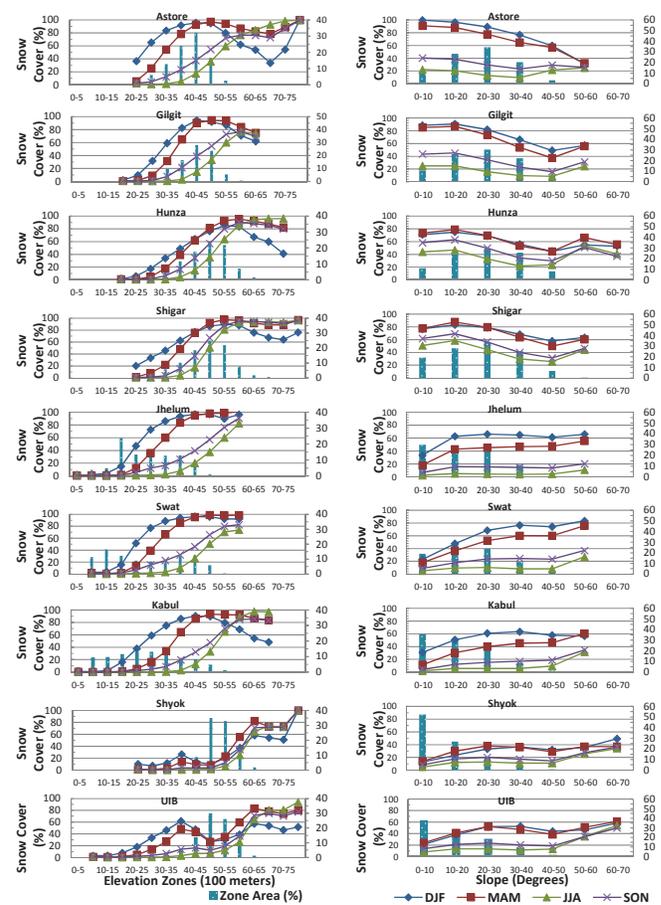

**Figure 11.** Snow cover against elevation (left panels) and slope (right panels) zones along with zone areas.

We find falling tendencies of the end-of-summer regional SLA zone for all basins except for the Astore sub-basin. Such tendencies are statistically significant for the Shyok and Kabul basins (Table 7). Furthermore, we find that the estimated SLA zones are situated well below the glaciers' median elevation within all basins (Fig. 12).

As for the snow cover relationship with slope, our findings suggest that Gilgit, Hunza, Astore and Shigar basins have experienced a low snow cover over higher slopes as accumulated snow cannot stay longer at steeper slopes. Instead, the Jhelum, Swat, Kabul, Shyok and UIB basins have experienced a low snow cover over lower slopes because their large surface areas extend to warmer, less mountainous regions (Fig. 11).

**Table 7.** Estimated end-of-summer mean regional SLA zone for each studied basin. Note: statistically significant trends at the 95 % level are shown in bold and italic, while trends significant at 90 % are shown in bold only. Negative values indicate a drop in the SLA zone, while positive values indicate its rise.

| S. no. | Basin | Mean regional SLA zone | SLA zone trend slopes |
| --- | --- | --- | --- |
| 1 | Astore | 4100–4200 | 17.13 |
| 2 | Gilgit | 3900–4000 | −9.79 |
| 3 | Hunza | 3400–3500 | −1.40 |
| 4 | Jhelum | 3600–3700 | −91.96 |
| 5 | Kabul | 3900–4000 | **−40.21** |
| 6 | Swat | 4500–4600 | −6.64 |
| 7 | Shigar | 3800–3900 | −0.70 |
| 8 | Shyok | 4200–4300 | ***−9.09*** |
| 9 | UIB | 3200–3300 | −20.63 |

### 5.3.5 Aspect-wise snow cover estimates

The quantitative snow cover dependence on aspect differs from location to location because, e.g., precipitation is impacted by the relationship between the aspect and the prevailing wind direction. Hence, we have calculated north-to-south (N–S), northwest-to-southeast (NW–SE), northeast-to-southwest (NE–SW) and west-to-east (W–E) ratios of seasonal snow cover for all the basins (Fig. 13 and Table 8). It is found that N–S ratios are high during all seasons, with the maximum during autumn. Such ratios are also higher than the other aspect ratios during all seasons. Only the Shigar and Shyok sub-basins experience a maximum N–S ratio during the winter season. Similarly, NE–SW ratios are high for all the basins and during all the seasons, except for Jhelum and Kabul during spring, and only for Jhelum during the summer season. Except for Astore in summer, all the other basins have either high or similar NW–SE ratios. The Shyok sub-basin and UIB experience low W–E ratios in the spring and winter seasons, whereas a similarly low ratio is found for the Hunza sub-basin only during the winter season. By combining the information contained in Table 8 and Fig. 13, we have derived the fact that the N–S, NE–SW and NW–SE ratios tend to be higher when the overall snow cover was lower. It points to the fact that aspect is not a very strong limiting factor for snow persistence in high, well snow-fed basins during colder seasons. The snowmelt due to direct sunlight becomes more relevant when climate conditions are milder and/or snow precipitation is weaker. Overall, the aspect-wise snow cover analysis shows that the Astore, Gilgit, Hunza, Shigar, Shyok, Jhelum and UIB basins have relatively larger basin areas at the northeastern and southwestern aspects, while they experience greater snow cover at the northeastern to northwestern aspects during all seasons.

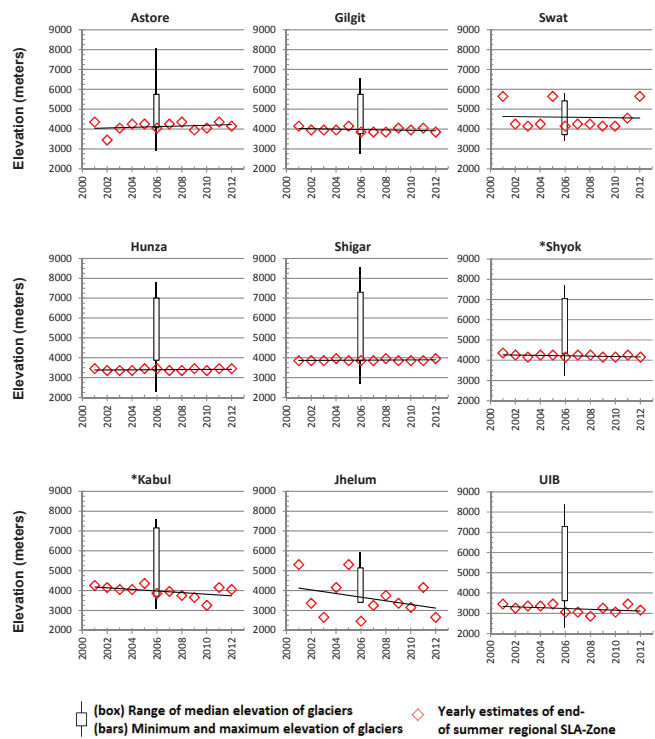

**Figure 12.** Median elevation of glaciers and inter-annual variability of the end-of-summer regional SLA zone for each studied basin. Note: basins marked with (*) feature statistically significant trends. The median elevation of glaciers along with the minimum and maximum elevations shown in the box plot are time independent.

### 5.4 Tele-connections

We have found a negative correlation between the autumn season (SON) NAO index and winter season snow cover for all the studied basins, which was particularly significant for UIB (Table 9). On the other hand, there was a significant positive correlation between the winter season (DJF) NAO index and the winter season (DJFM) snow cover for the Jhelum and Kabul basins. This may be due to the fact that local forcing such as topography (Cintia and Uvo, 2003; Bojariu and Giorgi, 2005), atmospheric aerosols (Cubasch et al., 2001), and additionally the complex interplay between the monsoon and the westerly disturbances during the summer seasons (Archer and Fowler, 2004), can all significantly alter the NAO effect.

There is a possibility that the strong connection (correlation coefficient greater than or equal to 0.3) found between a short-length snow cover time series and a large-scale phenomenon – showing variability on quasi-biennial and quasi-decadal timescales (Hurrel and Van Loon, 1997) – may result from the anomalous behaviour of the used snow product. We test this hypothesis against the long-term precipitation observations from PMD meteorological stations. We find, for the 1961–2000 period, a strong positive correlation of the long-term winter season precipitation with the NAO index, mainly for the November–January period. In Table 10, we show only those stations that feature statistically significant correlations of greater than or equal to 0.3 for at least one of the seasons.

**Table 8.** Seasonal aspect ratios of snow coverage for all the study basins.

| S. no. | Basin name | DJF | | | | MAM | | | | JJA | | | | SON | | | |
|---|---|---|---|---|---|---|---|---|---|---|---|---|---|---|---|---|---|
| | | N/S | W/E | NE/SW | NW/SE | N/S | W/E | NE/SW | NW/SE | N/S | W/E | NE/SW | NW/SE | N/S | W/E | NE/SW | NW/SE |
| 1 | Astore | 1.2 | 1.0 | 1.1 | 1.1 | 1.2 | 1.0 | 1.1 | 1.1 | 1.3 | 1.0 | 1.1 | 0.9 | 1.7 | 1.0 | 1.3 | 1.2 |
| 2 | Gilgit | 1.2 | 1.0 | 1.2 | 1.1 | 1.3 | 1.0 | 1.1 | 1.1 | 1.2 | 1.1 | 1.1 | 1.1 | 1.5 | 1.1 | 1.2 | 1.3 |
| 3 | Hunza | 1.3 | 0.9 | 1.3 | 1.1 | 1.3 | 1.0 | 1.2 | 1.1 | 1.4 | 1.0 | 1.2 | 1.1 | 1.4 | 1.0 | 1.3 | 1.2 |
| 4 | Jhelum | 1.2 | 1.1 | 1.0 | 1.1 | 1.3 | 1.1 | 0.9 | 1.2 | 1.5 | 1.3 | 0.9 | 1.4 | 1.9 | 1.2 | 1.1 | 1.7 |
| 5 | Kabul | 1.2 | 1.2 | 1.0 | 1.2 | 1.2 | 1.3 | 0.9 | 1.2 | 1.3 | 1.3 | 1.0 | 1.1 | 1.5 | 1.4 | 1.0 | 1.5 |
| 6 | Shigar | 1.1 | 1.0 | 1.2 | 1.1 | 1.1 | 1.0 | 1.1 | 1.1 | 1.0 | 1.0 | 1.1 | 1.0 | 1.1 | 1.0 | 1.2 | 1.1 |
| 7 | Shyok | 1.8 | 0.9 | 1.5 | 1.1 | 1.6 | 0.9 | 1.4 | 1.1 | 1.3 | 1.0 | 1.2 | 1.1 | 1.7 | 1.0 | 1.4 | 1.3 |
| 8 | Swat | 1.2 | 1.0 | 1.1 | 1.1 | 1.2 | 1.0 | 1.1 | 1.1 | 1.2 | 1.2 | 1.0 | 1.2 | 1.4 | 1.1 | 1.1 | 1.4 |
| 9 | UIB | 1.5 | 0.9 | 1.3 | 1.1 | 1.4 | 0.9 | 1.2 | 1.1 | 1.3 | 1.0 | 1.2 | 1.1 | 1.6 | 1.0 | 1.3 | 1.4 |

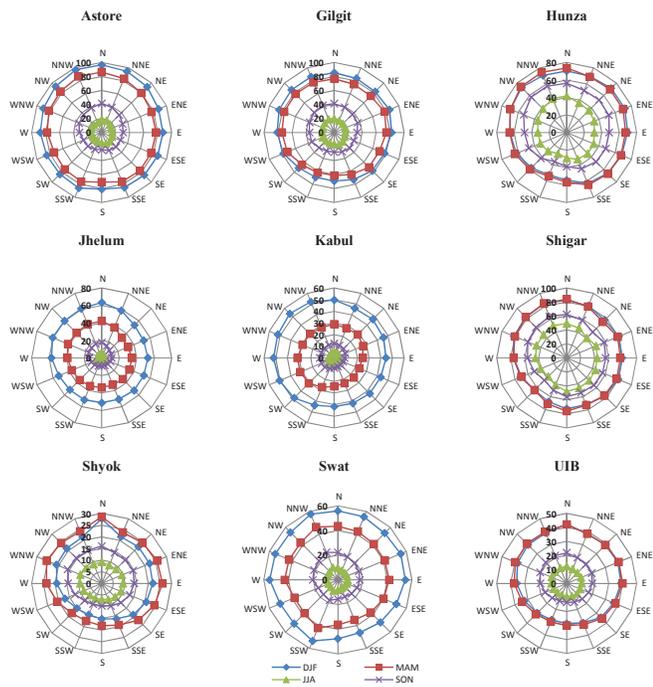

**Figure 13.** Aspect-wise distribution of snow cover (in percentage) during the winter (DJF), spring (MAM), summer (JJA) and autumn (SON) seasons for the nine study basins.

Our correlation analysis suggests a significant correlation of the spring season snow cover, spring season long-term precipitation and estimated winter/spring SWE with the summer season discharge at nearby or corresponding discharge stations of mainly the snow-fed basins (Tables 11–13). For the glacierised basins, we note that the summer season mean temperatures explain well the variability of summer season discharge (Table 11).

## 6 Discussion

Understanding of the statistical properties of snow cover, of its seasonal and inter-annual variability, and its slow trend dynamics is crucial for understanding the hydrological system and the role of snowmelt runoff within the Indus basin, and has important consequences for the study of its socio-ecological systems.

The MODIS instrument provides a good opportunity to assess the snow cover dynamics over the study region, mainly due to its high temporal resolution. The validation results of the MODIS snow products are found to be consistent with the previous studies performed over parts of the study region (Tahir et al., 2011a; Forsythe et al., 2012) and over the neighbouring regions within the Hindu Kush (Gafurov and Bárdossy, 2009) and the western Himalayan (Jain et al., 2008; Chelamallu et al., 2014) ranges, all suggesting that the use of the MODIS snow products is effective for the mapping of snow cover under Himalayan conditions.

Relative and absolute validations of our cloud removal technique suggest that the applied technique not only removes the clouds but also improves the overall accuracy of the used MODIS snow products against the observations. We note that the cloud removal technique is unable to completely remove the clouds from the snow products. Generally, this happens when clouds persist longer than the window size of a temporal filling or the cloud is larger in extent as compared to the window size of the applied spatial filling. Such conditions are observed over the study region during winter and spring seasons. Therefore, our cloud removal technique does not perform well during respective seasons, particularly for the high-latitude/altitude glacierised sub-basins (Hunza and Shigar) during winter season. Nevertheless, its overall satisfactory performance encourages its application to reduce/remove the cloud cover and improve snow cover. In view of the influence of cloud cover on the analysis (Hall et al., 2002), estimates of the snow cover-runoff relationship and of the climate change impacts on melt water runoff – based on the MODIS snow products not improved for clouds (Immerzeel et al., 2009; Tahir et al., 2011b; Forsythe et al., 2012) – may be prone to larger uncertainties.

Table 9. Correlations between the NAO index and the snow cover (SC) of the studied basins. Correlations significant at the 90 % level are marked in italic only, at the 95 % level are marked in bold only, and at the 99 % level are marked in bold and italic.

| S. no. | Basin | SCA(DJF)–NAO(DJF) | SCA(DJFM)–NAO(DJF) | SCA(DJF)–NAO(ASO) | SCA(DJF)–NAO(SON) | SCA(DJF)–NAO(OND) | SCA(DJF)–NAO(NDJ) |
|---|---|---|---|---|---|---|---|
| 1 | Astore at Doyian | 0.18 | 0.13 | −0.39 | **−0.62** | *−0.52* | −0.15 |
| 2 | Gilgit at Gilgit | 0.22 | 0.13 | −0.19 | −0.45 | *−0.58* | −0.26 |
| 3 | Hunza at Dainyor Bridge | −0.10 | −0.04 | −0.41 | **−0.66** | **−0.67** | −0.39 |
| 4 | Jhelum at Azad Patan | *0.54* | **0.70** | 0.07 | −0.18 | −0.15 | 0.06 |
| 5 | Kabul at Nowshera | *0.52* | **0.59** | 0.30 | −0.01 | −0.20 | −0.11 |
| 6 | Swat at Chakdara | 0.43 | 0.35 | −0.03 | −0.27 | −0.40 | −0.15 |
| 7 | Shigar at Shigar | 0.06 | 0.11 | −0.28 | −0.50 | −0.34 | −0.32 |
| 8 | Shyok at Yugo | −0.45 | −0.22 | *−0.59* | ***−0.74*** | −0.51 | 0.06 |
| 9 | UIB at Besham Qila | −0.33 | −0.25 | *−0.55* | ***−0.79*** | **−0.67** | −0.10 |

Table 10. Correlations between the NAO index and the long-term precipitation from PMD meteorological stations (1961–2000) within the studied basins. Only those stations which feature a statistically significant correlation of greater than or equal to 0.3 for at least one of the seasons are presented here. Correlations significant at the 90 % level are marked in italic, and at the 95 % level are marked in bold.

| S. no. | Lower-altitude stations (PMD) | Study basin | DJF–winter | DJF–DJF | DJF–ASO | DJF–SON | DJF–OND | DJF–NDJ |
|---|---|---|---|---|---|---|---|---|
| 1 | Astore | Astore | 0.26 | *0.27* | −0.06 | −0.04 | 0.13 | *0.31* |
| 2 | Garhi Dupatta | Jhelum | **0.33** | 0.25 | −0.04 | −0.07 | 0.10 | **0.35** |
| 3 | Muzaffarabad | Jhelum | **0.44** | **0.40** | −0.25 | −0.19 | 0.11 | **0.39** |
| 4 | Skardu | UIB | **0.32** | *0.30* | −0.21 | −0.16 | 0.13 | **0.35** |
| 5 | Chitral | Kabul | 0.14 | 0.20 | **−0.38** | **−0.39** | −0.06 | 0.22 |
| 6 | Jhelum | Jhelum | **0.32** | 0.17 | 0.07 | −0.08 | −0.06 | 0.10 |
| 7 | Murree | Jhelum | **0.41** | *0.31* | 0.00 | −0.05 | 0.11 | **0.35** |

Table 11. Correlation coefficients between the long-term seasonal precipitation (P) or average temperature (Tavg), with the seasonal discharge (Q). Note: values significant at the 95 % level are marked in bold.

| S. no. | Station | Basin | P (DJF)–Q (MAM) | P (MAM)–Q (JJA) | Tavg (JJA)–Q (JJA) |
|---|---|---|---|---|---|
| 1 | Muzaffarabd | Jhelum | **0.78** | 0.46 | |
| 2 | Azad Pattan | Jhelum | **0.56** | **0.73** | |
| 3 | Garhi Dupatta | Jhelum | 0.37 | **0.64** | |
| 4 | Balakot | Jhelum | 0.37 | **0.64** | |
| 5 | Peshawar | Kabul | 0.37 | **0.60** | |
| 6 | Saidu Sharif | Kabul | **0.64** | **0.65** | |
| 7 | Drosh | Kabul | **0.57** | **0.56** | |
| 8 | Chitral | Kabul | 0.56 | **0.63** | |
| 9 | Astore | Astore | 0.29 | **0.58** | |
| 10 | Gilgit | Gilgit | 0.09 | 0.35 | |
| 11 | Gupis | Gilgit | 0.28 | **0.53** | |
| 12 | Skardu | Shigar | 0.41 | 0.12 | 0.48 |
| 13 | Skardu | Shyok | −0.04 | 0.09 | **0.51** |
| 14 | Gilgit | Hunza | −0.16 | 0.11 | **0.36** |
| 15 | Skardu | UIB | 0.10 | *0.43* | |
| 16 | Saidu Sharif | Swat | 0.32 | 0.35 | |

**Table 12.** Correlation coefficients for the estimated SWE (accumulated precipitation during periods when temperatures remained below zero), with the summer discharge ($Q$) at the nearby or corresponding gauging stations. Note: values significant at the 95 % level are marked in bold, and at the 90 % level in italic.

| Basin name of the gauging stations | Burzil | Deosai | Hushe | Kelash | Khod | Khunjrab | Naltar | Rama | Rattu | Saiful Maluk | Shangla | Shendure | Shogran | Ushkore | Yasin | Zanipass | Ziarat |
|---|---|---|---|---|---|---|---|---|---|---|---|---|---|---|---|---|---|
| Astore  |     | **0.3** | **0.6** |     |     |      |      |      | 0.4  | −0.4 |     |     |      |     |     |     |     |
| Gilgit  |     |     |     |     |     |      |      |      |      |     |     |     |      | 0.2 | 0.2 |     |     |
| Hunza   |     |     |     |     |     | −0.1 | 0.0  |      |      |     |     |     |      |     | 0.2 |     | −0.2 |
| Jhelum  | **0.8** |     |     |     |     |      |      |      |      | 0.3 |     |     | **−0.5** |     |     |     |     |
| Kabul   |     |     |     | 0.5 | 0.4 |      |      |      |      |     | −0.1 | **0.9** |      |     |     | 0.1 |     |
| Shigar  |     |     |     |     |     |      |      |      |      |     |     |     |      |     |     |     |     |
| Shyok   |     | −0.2 | 0.1 |     |     |      |      | −0.4 | −0.2 |     |     |     |      |     |     |     |     |
| Swat    |     |     |     |     |     |      |      |      |      |     | −0.1 | 0.5 |      |     |     |     |     |
| UIB     |     | 0.2 | **0.6** |   |     | 0.2  | 0.1  | 0.1  | −0.4 |     |     |     |      | −0.1 | **0.6** |     | −0.2 |

A slight underestimation of the glacier area mapped in the Randolph Glacier Inventory (RGI – Pfeffer et al., 2014) by the MODIS minimum snow cover, for most of the studied basins, is mainly due to the inability of the MODIS sensor to detect debris-covered parts of the glaciers (Painter et al., 2012), which are quite common in the region (Bolch et al., 2012). Such underestimation is relatively pronounced for the Hunza sub-basin, which has substantial debris-covered glacier area. Additionally, discrimination between snow and ice in the MODIS binary products might be poorly achieved through automated global algorithm (Hall and Riggs, 2007; Shea et al., 2013) and may also result in such underestimation. On the other hand, Astore basin shows a large inconsistency between its MODIS minimum snow cover and the total glacier area. This may be attributed to the fact that the observation dates of the available glacier data used in RGI and of the extracted minimum MODIS snow cover do not necessarily coincide, plus the unavoidable issues associated to the quality of data in a complex terrain (Painter et al., 2012) of the HKH region. Coarser resolution of the MODIS products as compared to the data used in RGI may also be responsible for lower estimates of snow cover as compared to glacial extent.

We have found that basins under monsoon influence behave differently to basins under the westerlies influence. In the context of increasing winter precipitation (Archer and Fowler, 2004), decreasing mean annual snow cover for westerlies-influenced basins indicates (1) enhanced melting due to the observed warming during winter (Fowler and Archer 2006; Khattak et al., 2011), and (2) the transformation of solid precipitation into liquid precipitation (Hasson et al., 2014). On seasonal timescales, snow cover decreases during the winter and autumn seasons, but increases during the summer and spring seasons, in agreement with Immerzeel et al. (2009) and with reports of cooling and warming trends during the respective seasons (Fowler and Archer, 2006; Gioli et al., 2014). It is pertinent to mention here that since we do not know the water content of the snowpack, snow cover increase or decrease does not necessarily correspond to an increase or a decrease in the water resources. However, such variations certainly contribute to the snowmelt runoff variability on spatio–temporal scales.

**Table 13.** Correlation coefficients between the seasonal snow cover area (SCA) for spring (MAM) and winter (DJF) with the discharge (Q) of the summer season (JJA). Note: values significant at the 95% level are marked in bold, and at the 90% level in italic. The Shigar gauging site is no longer operational, so far.

| S. no. | Basin | SCA(DJF)–Q(MAM) | SCA(MAM)–Q(JJA) |
|---|---|---|---|
| 1. | Atore  | **0.56** | **0.70** |
| 2. | Gilgit | −0.18 | **0.59** |
| 3. | Hunza  | −0.42 | *−0.49* |
| 4. | Shigar | – | – |
| 5. | Shyok  | −0.35 | **0.54** |
| 6. | UIB    | 0.01  | **0.47** |
| 7. | Jhelum | −0.05 | **0.57** |
| 8. | Kabul  | −0.06 | 0.30 |
| 9. | Swat   | **0.52** | **0.66** |

Here, we extend the general picture of precipitation distribution presented in section 5.3 by summarizing reports from different studies. Winiger et al. (2005) have reported the snow depths of around 1200 and 1800 mm at Dame (36°01′ N, 74°35′ E at 3670 m a.s.l.) and Diran (36°04′ N, 74°36″ E at 4050 m a.s.l.) stations, respectively, in Bagrot Valley – 20 km northeast of Gilgit, Karakoram range. They also reported that, along the Gilgit-Khunjrab transect within the Hunza basin, precipitation ranges between 600 and 1200 mm within the altitude belt of 3500–4500 m a.s.l., of which 90 % is incident as snow, while it is around 400 mm below 3000 m a.s.l., occurring only 10 % as snow. In the central Karakoram, Hewitt (2011) has reported the maximum precipitation at 4800 m a.s.l., which occurs entirely in solid form. At 4840 m a.s.l., Batura Investigation Group (1979) had reported the annual snowfall around 1034 mm over the Batura Glacier lying in the eastern part of the Hunza sub-basin. For the altitudinal belt of 4800–5800 m a.s.l. over the Biafo and Hispar glaciers (along the transect of Hunza and Shigar sub-basins), Wake (1987) reported that the snowfall here generally exceeds 1000 mm and at some locations it exceeds 2000 mm water equivalent.

Relative to normal conditions, a fall in the end-of-summer regional SLA zone features fewer glacier/ice/snow surface areas exposed to melting at higher elevations, indicating a two-fold positive change: (1) the glaciers lose relatively less mass, and (2) the glacier/ice mass increases because of the remaining accumulated snow. On the other hand, a rise in the end-of-summer regional SLA zone features increased exposure of snow and snow-free glacier areas to thaw conditions at higher elevations, indicating enhanced melting. In view of the absence of snowpack water content information, we do not know how much moisture is accumulated due to the observed descent of the SLA zone; however, such conditions confirm at least less melting at higher elevations, indicating a positive change in the water resources therein. The placement of a regional SLA zone well below the median elevation of existing glaciers – a proxy of ELA (Braithwaite and Raper, 2009) – also confirms an indication of a positive mass balance of these glaciers. Such a proxy finding is further confirmed by Gardelle et al. (2013), who show a possible slightly positive mass balance of the Karakoram glaciers for the last decade, by Fowler and Archer (2006), showing observed cooling of the summer season, and by Khattak et al. (2011), reporting a subsequent reduction in the summer season flows.

As meltwater largely contributes to the overall freshwater availability in Pakistan (Immerzeel et al., 2010; Hasson et al., 2014), less summer melt resulting from the above-described situation leads to decreased water availability downstream, indicating an alarming situation for the water resource management in the country. The findings based on modelling studies, that initially the stream flow will increase due to increased melting and then will abruptly decrease when glaciers disappear (Rees and Collins, 2006; Akhtar et al., 2008; Immerzeel et al., 2009; Tahir et al., 2011b), are quite misleading at present for many reasons. These include the fact that (1) the region has not yet completely followed the warming signal of climate change as observed globally or as projected by the present-day climate models (increase in temperature), (2) climate models have generally been unable to describe such anomalies over the region (Hasson et al., 2013, 2014), and (3) it is not known when the summer cooling phenomena will come to an end. Most commonly adopted scenarios of temperature increase and glacier area decrease, therefore, presently seem inappropriate for short-term scenarios (see Fowler and Archer, 2006, Khattak et al., 2011, and Minora et al., 2013), but may be relevant for the long-term future. Under such a scenario, we encourage the modelling community to consider additionally the observed hydro-climatic scenario, in order to assess the near-future melt-runoff contribution to the hydrology of the region.

The significant correlation found for snow cover and long-term precipitation with the NAO index suggests the forecast possibility of these variables with a lead time of one month up to a season. Archer and Fowler (2004) also found a strong significant correlation between winter precipitation and the November-to-January NAO index for a few stations considered in the study. The strong correlation found between the stream flow observations and precipitation suggests good potential for providing indicators of runoff variability during the melt season in the region. Based on such a relationship, Archer and Fowler (2008) performed a statistically based seasonal stream flow forecast for the Jhelum basin using long-term precipitation data. Similarly, the spring season snow cover shows a strong correlation with the summer season discharge of the corresponding basins. Forsyth et al. (2012) also reported a similar connection for the Astore basin. Though short in length, seasonal snow cover shows a similar strength of correlation between NAO and stream flow as per the long-term precipitation record. For the south Asian summer monsoonal precipitation, Immerzeel and Bierkens (2010) showed that spring season snow cover over the Tibetan Plateau can serve as an important predictor, when combined with the global indices of ocean–atmospheric modes such as NAO and ENSO. Therefore, our findings here suggest the possibility of a statistically based stream flow forecast for the region in advance – or at least an opportunity to address the stream flow variability – allowing snow cover as one of the predictors.

## 7 Conclusions

This study constitutes an effort to understand the present state of a snow-cover regime and its dynamics and its temporal and spatial variability in the region, taking into account different geophysical parameters. The data set time frame is too short to allow for robust conclusions about the general behaviour and long-term changes of snow cover. However, observed tendencies of snow cover and of the SLA zone further confirm the trends in related variables as reported by different studies (Archer and Fowler, 2004; Khattak et al., 2011; Gardelle et al., 2013; Gioli et al., 2014). The main findings of our study are summarised here:

– The westerlies-influenced basins (UIB, Hunza, Shigar, Shyok and Astore) show a decreasing snow cover tendency, while the monsoon-influenced basins (Jhelum, Swat, Kabul and Gilgit) show an increasing snow cover tendency on an annual timescale. All basins show decreasing snow cover trends during the winter and autumn seasons, except for Gilgit during the autumn season. The Jhelum basin shows an increasing snow cover trend throughout the year, while the rest of the basins show increasing snow cover trends for the spring and summer seasons (except Shyok and UIB). However, only the summer snow cover trend for Jhelum was statistically significant.

– High variability found for the snow accumulation and ablation seasons is relatively more pronounced for the snow-fed than for the glacier-fed basins, and during the winter and spring seasons than during the summer and autumn seasons. An east–west gradient is not present in terms of snow cover and its variability. However, sub-basins at higher latitudes/altitudes show more snow cover variability than basins at relatively lower latitudes/mid-altitudes.

– High seasonal variation with respect to elevation is observed for the snow-fed basins as compared to the glacier-fed basins throughout the year. The average regional SLA zone for the Hunza and UIB basins ranged from 3000 to 3500 m a.s.l., for the Gilgit, Shigar, Jhelum and Kabul basins from 3500 to 4000 m a.s.l., for the Astore and Shyok basins from 4000 to 4500 m a.s.l., and for the Swat basin from 4500 to 5000 m a.s.l.

– The Astore, Gilgit, Hunza, Shigar, Shyok, Jhelum and UIB basins have comparatively larger areas at the northeastern and southwestern aspects, and greater snow cover at the northeastern aspect. Northern aspects, as expected, have in general more snow cover than southern aspects. Such discrepancies become larger when considering warmer seasons or basins at lower altitudes, where temperature is a strong limiting factor for the snow persistence.

– The Gilgit, Hunza, Astore and Shigar basins have low snow cover over higher slopes, whereas the Jhelum, Swat, Kabul, UIB and Shyok basins have low snow cover over lower slopes.

– Under the prevailing climatic conditions, there was an indication of a positive change in the frozen water resources of the region, particularly for the UIB. This was evident from the facts of (1) an observed increase in winter and summer precipitation (Archer and Fowler, 2004), (2) a possible positive mass balance of the central Karakoram glaciers (Gardelle et al., 2013), (3) decreasing summer flows (Khattak et al., 2011), (4) a falling end-of-summer regional SLA zone (significant for the Shyok and Kabul basins), and (5) increasing summer season snow cover. In contrast, a warming trend during the winter season and a consequent increase in the winter season flows (Khattak et al., 2011) indicated a possible seasonal shift in the snow distribution. Such a shift depends partly on the precipitation input during the accumulation season and partly on the prevailing seasonal temperature regimes. The observed trends agreed with the recently collected local perceptions of the climate change and variability (Gioli et al., 2014).

– A significant correlation of snow cover and precipitation with the NAO index, and furthermore with the stream flow, reveals the possibility of a short-term forecast of water resources with a lead time from one month up to a season, suggesting snow cover as one of the predictors.

**The Supplement related to this article is available online at doi:10.5194/hess-18-4077-2014-supplement.**

*Acknowledgements.* The authors acknowledge the National Snow and Ice Data Center (NSIDC) and the MODIS team for their roles in making available the MODIS daily (Aqua and Terra) snow cover products for the period 2001–2012. S. Hasson acknowledges the support of BMBF, Germany's CLASH/Climate variability bundle project and landscape dynamics in southeastern Tibet and the eastern Himalayas during the late Holocene reconstructed from tree rings, soils and climate modelling. V. Lucarini acknowledges the support of the FP7/ERC Starting Investigator grant NAMASTE/Thermodynamics of the Climate System (grant no. 257106). The support from CliSAP/Cluster of excellence in the Integrated Climate System Analysis and Prediction is also acknowledged.

Edited by: H.-J. Hendricks Franssen